\documentclass[11pt,preprint,superscriptaddress,amsmath,amssymb,11pt,aps,showpacs,showkeys,nofootinbib,a4paper]{article}
\usepackage{amssymb,amsmath,amsfonts}
\usepackage{graphicx}
\usepackage{eepic,epsfig}
\usepackage{verbatim}
\usepackage{color}
\usepackage{hyperref}
1.60cm \makeatletter \@addtoreset{equation}{section}

\makeatother

\begin{document}
\title{Scalar Casimir effect in a high-dimensional cosmic dispiration spacetime}
\author{H. F. Mota\thanks{E-mail: hmota@fisica.ufpb.br}, E. R. Bezerra de Mello\thanks
{E-mail: emello@fisica.ufpb.br} and K. Bakke\thanks{E-mail: kbakke@fisica.ufpb.br}\\
\\
\textit{Departamento de F\'{\i}sica, Universidade Federal da Para\'{\i}ba}\\
\textit{58059-900, Caixa Postal 5008, Jo\~{a}o Pessoa, PB, Brazil}\vspace{%
0.3cm}\\
}
\maketitle
%
\begin{abstract}
In this paper we present a complete and  detailed analysis of the calculation of both the Wightman function and the vacuum expectation value of the energy-momentum tensor that arise from quantum vacuum fluctuations of massive and massless scalar fields in the cosmic dispiration spacetime, which is formed by the combination of two topological defects: a cosmic string and a screw dislocation. This spacetime is obtained in the framework of the Einstein-Cartan theory of gravity and is considered to be a chiral space-like cosmic string. For completeness we perform the calculation in a high-dimensional spacetime, with flat extra dimensions. We found closed expressions for the the energy-momentum tensor and, in particular, in (3+1)-dimensions, we compare our results with existing previous ones in the literature for the massless scalar field case. 

\end{abstract}
\bigskip
%
\section{Introduction}
\label{Int}
%

\hspace{0.55cm}Casimir effect is a macroscopic quantum manifestation of vacuum fluctuations by virtue of the imposition of boundary conditions on relativistic fields, namely, scalar, spinor and electromagnetic fields \cite{Mostepanenko:1997sw,bordag2009advances,Milton:2001yy}. This effect was first predicted in 1948 by Casimir \cite{Casimir:1948dh} who investigated the quantum vacuum fluctuations of the electromagnetic field occurring inside two large parallel conducting plates. His result was first confirmed by Sparnaay in 1958 \cite{Sparnaay:1958wg} and later by other researchers \cite{Bressi:2002fr} (see also \cite{Lamoreaux:1996wh, PhysRevLett.81.5475, mohideen1998precision, MOSTEPANENKO2000, PhysRevA.78.020101, PhysRevA.81.052115} for experimental results involving curved-plane surface configurations). It is well known that the modification on the quantum vacuum fluctuations of fields are caused not only by the imposition of boundary conditions but also by the nontrivial topology of the spacetime  \cite{Mostepanenko:1997sw,bordag2009advances}. Moreover, the current status of the Casimir effect establishes its reality and motivates the study of quantum vacuum fluctuations of other observables rather than only the Casimir energy density. 

In this direction, two types of spacetimes with nontrivial topology have mainly been considered in the literature, that is, screw dislocation (which is associated with torsion) and cosmic string. The latter is a linear topological defect supposed to be formed in the early Universe after it went through phase transitions and is predicted in the context of extensions of the Standard Model of particle physics \cite{VS, hindmarsh} as well as in the context of string theory \cite{Copeland:2011dx,Hindmarsh:2011qj}. Once formed, cosmic strings can evolve in the Universe and contribute to a variety of astrophysical, cosmological and gravitational phenomena  \cite{Copeland:2011dx,Hindmarsh:2011qj, Mota:2014uka}. From the static point of view, the presence of a idealized thin, straight and very long cosmic string produces a spacetime with a conical topology with a planar angle deficit on the two-surface orthogonal to it given by $\Delta\varphi = 8\pi G\mu_0$, where $G$ is the Newton's gravitational constant and $\mu_0$ the cosmic string linear energy density \cite{VS, hindmarsh}.

The investigation of quantum vacuum fluctuations of physical observables, such as the energy-momentum tensor, due to the conical topology of the cosmic string spacetime can be found in Refs.  \cite{BezerradeMello:2006df, PhysRevD.35.536, escidoc:153364, GL, DS, PhysRevD.46.1616, PhysRevD.35.3779, LB, Moreira1995365, BK,Aliev:1998qm}. Another physical observable of great interest is the induced quadri-current by quantum vacuum fluctuations associated to charged matter fields and which has been investigated in Refs. \cite{Braganca:2014qma,LS, SNDV, ERBM, PhysRevD.82.085033}. In the context of a linearized model of lightcone fluctuations \cite{Ford:1994cr}, a delay or advance in time in the propagation of photons arises as  a consequence of quantum vacuum fluctuations due to the conical topology of the cosmic string spacetime. The analysis of this observable, i.e, the shift in time, was considered in Ref. \cite{Mota:2016mhe}. 

Concerning the screw dislocation,  it is also considered to be a line-like topological defect found in the context of theories of solid and crystal continuum media and seen as a type of Volterra distorsion \cite{Puntigam:1996vy}.  In the framework of the Einstein-Cartan theory of gravity, on the other hand, the merger of a screw dislocation with a cosmic string is naturally seen as a topological defect having a chiral nature as well as a space-like helical structure, with delta function singularities present in both the scalar curvature and torsion, charactering the geometry of the spacetime \cite{Galtsov:1993ne,Letelier:1995ze}. Note that although the authors in Ref. \cite{Galtsov:1993ne} used the terminology `cosmic dislocation' to refer to this chiral cosmic string, we will follow the same terminology coined by the authors in Ref. \cite{DeLorenci:2002jv} who called this combination as `cosmic dispiration'. It seems more appropriate due to the singularity associated to the screw dislocation aspect of the defect. Also, the spacetime produced by a cylindrically symmetric gravitational wave has a large-distance asymptotic behaviour with the same structure as the spacetime generated by a cosmic dispiration \cite{Berger:1994sm, tod1994conical}.

In fact, the presence of a torsion in the spacetime has attracted a great deal of attention, for instance, due to interaction with fermions \cite{Audretsch:1981xn,Ryder:1998is, Shapiro:2001rz,Singh, Hagen}. A line of research dealing with torsion as a topological defect in crystalline solids with the use of differential geometry can be found in Refs. \cite{kleinert1989gauge,Katanaev:1992kh}. Examples of these defects associated with torsion are the screw dislocations and the spiral dislocations \cite{valanis2005material}. Several studies have explored the influence of these kind of topological defects on quantum systems, such as the quantum scattering \cite{furtado2001quantum}, the Aharonov-Bohm effect for bound states \cite{de2005quantum}, geometric quantum phases \cite{bakke2013abelian} and geometric quantum computation \cite{bakke2013one}. The influence of torsion has also attracted attention in condensed matter system described by the Dirac equation, for instance, graphene \cite{Carpio,mesaros,deJuan:2010zz}. Investigations about spacetimes generated by chiral cosmic strings, obtained as a solution of the Einstein-Cartan equations, can be found in Refs.  \cite{Galtsov:1993ne,Letelier:1995ze,Puntigam:1996vy,Bezerra:1997mn}. In this sense, torsion effects on relativistic quantum systems associated with the presence of these chiral cosmic strings have also been reported in the literature. Among them, it is worth citing the Aharonov-Bohm effect for bound states \cite{Bezerra:1997mn} and the studies of the Dirac oscillator \cite{Bakke:2013wla} and a relativistic position-dependent mass system \cite{Vitoria:2016bwq}.   

In Ref. \cite{Galtsov:1993ne}, the authors pointed it out that a cosmic dispiration produces lensing effects both in its transversal and longitudinal directions, in contrast with the cosmic string case where only the lensing effect in its transversal direction exists. In this sense, it is possible to distinguish the usual cosmic string from the chiral space-like cosmic string, that is, the cosmic dispiration.

In particular, only a few studies about cosmic dispiration spacetime, in the framework of quantum vacuum fluctuations, have been reported in the literature \cite{pontual1998casimir,DeLorenci:2002jv,DeLorenci:2003wv}. In Ref. \cite{pontual1998casimir} the authors performed the investigation of the scalar Casimir energy density in the spacetime of a screw dislocation using the zeta function technique and obtained only a long-range approximated expression. On the other hand, the authors in Ref. \cite{DeLorenci:2002jv} developed an investigation of quantum fluctuations of the massless scalar field in the spacetime of a cosmic dispiration and calculated the propagator and an approximated expression for the vacuum expectation value (VEV) of the energy momentum tensor. In this paper, we present an exact expression for the VEV of the energy-momentum tensor arising from quantum vacuum fluctuations of both the massive and massless scalar fields in a high-dimensional cosmic dispiration spacetime.

This paper is organized as follows. In Section 2 we introduce the spacetime of a cosmic dispiration and calculate the two-point function, namely, the Wightman function. Section 3 is devoted for the complete calculation of the expressions for the VEV of the field squared in both cases, the massive and massless scalar fields. Using results from the two previous Sections, in Section 4 we calculate the VEV of the energy-momentum tensor. Finally, in Section 5 we present our final remarks.  We have also dedicated an Appendix to obtain important expressions used in the development of our calculation of both the Wightman function and the VEV of the energy-momentum tensor. Throughout the paper we use natural units $G=\hbar=c=1$.

\section{Klein-Gordon equation and Wightman function}
\label{sec2}
In this section, we consider a spacetime with nontrivial topology that arises due to the combined effects of a cosmic string and a screw dislocation, that is, the spacetime of a cosmic dispiration. The solution of the Klein-Gordon equation is taken only assuming that the radial solution is regular at the origin. This solution allows us to investigate the effect of the nontrivial topology of the cosmic dispiration spacetime on the VEV of physical observables. 

Let us start by considering the following metric that describes an idealized ($D+1$)-dimensional cosmic dispiration spacetime in cylindrical coordinates:
\begin{equation}
ds^2=g_{\mu\nu}dx^{\mu}dx^{\nu}=dt^2-dr^2-r^2d\varphi^2-(dz+\kappa d\varphi)^2-\sum_{i=4}^{D}(dx^i)^2,
\label{Me}
\end{equation}
where $\kappa$ is a constant parameter associated with the screw dislocation\footnote{The constant $\kappa$ is actually called burgers vector in the framework of theories of continuum media \cite{Puntigam:1996vy}. In the context of the Einstein-Cartan theory of gravity it is identified with $\frac{2GJ^{z}}{\pi}$, where $J^{z}$ is a component of a two-dimensional vector tangent to the string worldsheet (see Refs. \cite{Galtsov:1993ne,Letelier:1995ze}  for more details). }, $D\geq3$ and $(r,\phi,z,x^4,...,x^D)$ are the generalized cylindrical coordinates taking values in the ranges $r\geq 0$, $0\leq\varphi\leq \varphi_0=2\pi/q$ and $-\infty<(t,z,x^i)<+\infty$, for $i=4,...,D$. The parameter $q\ge 1$ codifies the presence of the cosmic string which we assume to be on the $(D-2)$-dimensional hypersurface $r=0$. This parameter, in $D=3$, is associated with the linear mass density of the string, $\mu_0$, through the relation $q^{-1}=1-4G\mu_0$, where $G$ is the Newton's gravitational constant. The usual identification condition for the spatial points are $(r,\varphi,z,x^4,...,x^D)$ $\rightarrow$ $(r,\varphi+2\pi/q,z,x^4,...,x^D)$. 

The line element (\ref{Me}) is locally flat and can be written in a Minkowskian form if we define the new spatial coordinate $Z=z+\kappa\varphi$, providing 
\begin{equation}
ds^2=g_{\mu\nu}dx^{\mu}dx^{\nu}=dt^2-dr^2-r^2d\varphi^2-dZ^2-\sum_{i=4}^{D}(dx^i)^2,
\label{Me2}
\end{equation}
where the new identification condition is found to be
\begin{equation}
(r,\varphi,Z,x^4,...,x^D) \rightarrow (r,\varphi+2\pi/q,Z+2\pi\kappa/q,x^4,...,x^D).
\label{IC}
\end{equation}
Although the physical dimension of the spacetime is for $D=3$, we decided to extend our analysis for $D\geq 3$; so, in this way, we can also investigate how the results obtained depend on the extra dimensions. However, our truly interest is in the usual $(3+1)$ physical spacetime dimensions, which we will mostly focus our analysis.

The action associated with a scalar field in a curved $(D+1)$-dimensional spacetime is \cite{birrell1984quantum}:
\begin{eqnarray}
S=\frac12\int \ dx^{D+1}\sqrt{|g|}\left[g^{\mu\nu}\nabla_\mu\Phi\nabla_\nu\Phi-\left(m^2+\xi R\right)\Phi^2\right] \  ,
\label{action}
\end{eqnarray}
where $g={\rm det}(g_{\mu\nu})$, $R(x)$ is the Ricci scalar and $\xi$ an arbitrary numerical factor named curvature coupling. In the case of $m=0$, the action and hence the field equations are invariant under conformal transformation, if $\xi=\frac14\frac{D-1}{D}$. So, if one is to consider a nonminimally coupled scalar field one can promptly construct a conformal invariant formalism.


The Klein-Gordon equation for a nonminimally coupled scalar field, $\Phi(x)$, can be written in the form
\begin{eqnarray}
\left[\frac{1}{\sqrt{|g|}}\partial_{\mu}\left(\sqrt{|g|}g^{\mu\nu}\partial_{\nu}\right)+m^2+\xi R\right]\Phi(x)=0,
\label{KG}
\end{eqnarray}
where $x$ in the argument of $\Phi(x)$ represents the set of all the spacetime coordinates.

Because we are considering an idealized cosmic dispiration spacetime, the scalar Ricci tensor, $R$, is a Dirac delta-type function, and only vanishes  for $r\neq 0$. So, all the wave-functions that do not vanish at origin present a direct interaction with the curvature coupling parameter. In order to avoid a hard and long analysis that is out of our main objective in this paper, we shall adopt $\xi=0$.\footnote{The analysis of the  Klein-Gordon equation in a pure cosmic string spacetime considering the presence of the curvature coupling, $\xi$, was developed by J. Spinelly at al \cite{Spinally:2000ii}. In the latter it was shown that for Bessel functions of order equal to zero, the analysis of the scattering problem becomes more delicate, and  the curvature coupling parameter must be renormalized.}

Using the line element (\ref{Me}), we can further write Eq. (\ref{KG}), for $\xi=0,$ in the form 
\begin{eqnarray}
\left[\frac{\partial^2}{\partial t^2}-\frac{1}{r}\frac{\partial}{\partial r}\left(r\frac{\partial}{\partial r}\right)-\frac{1}{r^2}\left(\frac{\partial}{\partial\varphi}-\kappa\frac{\partial}{\partial z}\right)^2-\frac{\partial^2}{\partial z^2}-\sum_{i=4}^{D}\frac{\partial^2}{\partial x^{i2}}+m^2\right]\Phi(x)
=0.
\label{KG2}
\end{eqnarray}
Since the Hamiltonian associated with the system commutes with the projection of the angular momentum operator along the $z$-axis, $L_\phi=-i\partial_\phi$, and with the linear momenta operators, $p_j=-i\partial_j$, for $j=3, ... , D$, we can adopt the {\it ansatz} below to obtain a solution of the partial differential equation \eqref{KG2}:
\begin{eqnarray}
\Phi_k(x)=CR(r)e^{-i\omega_kt+inq\varphi+i\nu z+i{\bf p}\cdot {\bf r_{\parallel}} },
\label{SS}
\end{eqnarray}
where $C$ is a normalization constant and ${\bf r_{\parallel}}$ and ${\bf p}$ represent, respectively, the coordinates of the extra dimensions and their corresponding momenta. Substituting Eq. (\ref{SS}) into (\ref{KG2}) we get the following differential equation
\begin{eqnarray}
\left[\frac{d^2}{dr^2}+\frac{1}{r}\frac{d}{dr}+\eta^2-\frac{(qn-\kappa\nu)^2}{r^2}\right]R(r)=0,
\label{KG3}
\end{eqnarray}
where $\omega_k^2=m^2+\eta^2+\nu^2+p^2$ and $k=(\eta,n,\nu,p)$ is the set of quantum numbers. The above equation is a Bessel differential equation whose solution is a combination of the Bessel $J_{\mu}(x)$ and Neumann $N_{\mu}(x)$ functions. Since the order of these functions depends on the continuum quantum number $\nu$, the Neumann function, in general, cannot be square-integrable at origin. Thus, we have to consider only the regular solution at origin of Eq.  (\ref{KG3}), that is,
\begin{eqnarray}
R(r)=J_{|qn-\kappa\nu|}(\eta r).
\label{BS}
\end{eqnarray}

The normalization constant $C$ can be calculated using the orthonormalization condition
\begin{eqnarray}
\int d^{D}x\sqrt{|g|}\Phi_k(x)\Phi^{*}_{k'}(x)=\frac{1}{2\omega_k}\delta_{k,k'},
\label{OD}
\end{eqnarray}
where the delta symbol on the right-hand side is understood as Kronecker delta for $n$ and Dirac delta function for the continuous quantum numbers, $\eta$, $\nu$ and ${\bf p}$. Using Eqs. (\ref{SS}) and (\ref{BS}), the orthonormalization condition in Eq. (\ref{OD}) provides
\begin{eqnarray}
|C|^2=\frac{q\eta}{2\omega_k(2\pi)^{D-1}}.
\label{OD2}
\end{eqnarray}
Therefore, the final form for the solution (\ref{SS}) is given by
 \begin{eqnarray}
\Phi_k(x)=\left[\frac{q\eta}{2\omega_k(2\pi)^{D-1}} \right]^{\frac{1}{2}}J_{|qn-\kappa\nu|}(\eta r)e^{-i\omega_kt+inq\varphi+i\nu z+i{\bf p}\cdot {\bf r_{\parallel}} }.
\label{SS2}
\end{eqnarray}
Equivalently, in terms of the the coordinates $(t,r,\varphi,Z,x^4,...,x^D)$, the general solution obeying the identification condition (\ref{IC}) is given by
 \begin{eqnarray}
\Phi_k(x)=\left[\frac{q\eta}{2\omega_k(2\pi)^{D-1}} \right]^{\frac{1}{2}}J_{|qn-\kappa\nu|}(\eta r)e^{-i\omega_kt+i(nq-\kappa\nu)\varphi+i\nu Z+i{\bf p}\cdot {\bf r_{\parallel}} }.
\label{SS3}
\end{eqnarray}
Note that although the solutions  (\ref{SS2}) and  (\ref{SS3}) lead to the same physical results, it is more convenient to work with the latter one since the line element (\ref{Me2}) allows us, as we will see, to write the energy-momentum tensor in a simpler form.

The topology of the spacetime described by the line element (\ref{Me2}) along with the requirement that the radial solution (\ref{BS}) be regular at the origin change the spectrum of vacuum fluctuations when compared with the Minkowski spacetime. Thus, in order to study the corresponding changes in the VEV of physical observables it is necessary to describe the properties of the vacuum state, $|0\rangle$, through the positive frequency Wightman function, $W(x,x')=\langle 0|\Phi(x)\Phi(x')|0\rangle$, being $\Phi(x)$ the field operator. The latter can be expanded in terms of the complete set of normalized mode functions, $\left\{\Phi_k(x),\Phi_k^{*}(x)\right\}$, given by Eq. (\ref{SS2})-(\ref{SS3}). Consequently, we obtain 
\begin{equation}
W(x,x')=\sum_{k}\Phi_k(x)\Phi_k^{*}(x'),
\label{WF}
\end{equation}
with the compact notation 
\begin{equation}
\sum_k=\int dp^{(D-3)}\int_{-\infty}^{\infty}  d\nu\int_{0}^{\infty}  d\eta\sum_{n=-\infty}^{\infty}. 
 \label{SI}
\end{equation}
The detailed calculation for the Wightman function is performed in the Appendix \ref{appA} and  the final result is given by
\begin{eqnarray}
W(x,x')&=&\frac{m^{(D-1)}}{(2\pi)^{\frac{(D+1)}{2}}}\left[\sum_{l} f_{\frac{D-1}{2}}\left(m\sigma_l\right)\right.\nonumber\\
&-&\left.\frac{q}{\pi^2}\sum_{n=-\infty}^{\infty}\int_{0}^{\infty}dy f_{\frac{D-1}{2}}\left(m\sigma_{y,n}\right)M_{n,q}(\Delta\varphi,y)\right],
\label{P32}
\end{eqnarray}
where the discrete index $l$ obeys the condition 
\begin{eqnarray}
-\frac{q}{2} + \frac{\Delta\varphi}{\varphi_0}\leq l \leq \frac{q}{2} + \frac{\Delta\varphi}{\varphi_0},
\label{A5text}
\end{eqnarray}
and we have used, in terms of the Macdonald function $K_{\mu}(x)$, the notation
\begin{equation}
f_{\mu}(x)=\frac{K_{\mu}(x)}{x^{\mu}}.
\label{No2}
\end{equation}
Note also that  
\begin{eqnarray}
\sigma_l^2 &=& \left[\Delta\zeta^2 - 2rr'\cos\left(2\pi l/q - \Delta\varphi\right) + \left(\Delta Z - \bar{\kappa}l\right)^2\right],\nonumber\\
\sigma_{y,n}^2 &=& \left[\Delta\zeta^2 + 2rr'\cosh(y) + \left(\Delta Z + \bar{\kappa}n\right)^2\right],
\label{GD2}
\end{eqnarray}
with $\Delta\zeta^2=\Delta\tau^2+\Delta {\bf r}_{\parallel}^2+r^2+r'^2$ and $\bar{\kappa}=\kappa/\varphi_0$. Additionally, the function $M_{n,q}$ is also presented in Appendix \ref{appA} and is written here as
\begin{eqnarray}
M_{n,q}(\Delta\varphi, y) = \frac{1}{2}\left\{\frac{\left(\frac{q}{2} + \frac{\Delta\varphi}{\varphi_0} + n\right)}{\left(\frac{q}{2} + \frac{\Delta\varphi}{\varphi_0} + n\right)^2 + \left(\frac{y}{\varphi_0}\right)^2} - \frac{\left(-\frac{q}{2} + \frac{\Delta\varphi}{\varphi_0} + n\right)}{\left(-\frac{q}{2} + \frac{\Delta\varphi}{\varphi_0} + n\right)^2 + \left(\frac{y}{\varphi_0}\right)^2}\right\}.
\label{A162}
\end{eqnarray}

On the other hand, by taking the limit $m\rightarrow 0$ in (\ref{P32}), the Wightman function for the massless case is found to be
\begin{eqnarray}
W(x,x')&=&\frac{2^{\frac{(D-1)}{2}}\Gamma\left(\frac{D-1}{2}\right)}{2(2\pi)^{\frac{(D+1)}{2}}}\left[\sum_{l} \frac{1}{\sigma_l^{(D-1)}}\right.\nonumber\\
&-&\left.\frac{q}{\pi^2}\sum_{n=-\infty}^{\infty}\int_{0}^{\infty}dy\frac{1}{\sigma_{y,n}^{(D-1)}}M_{n,q}(\Delta\varphi,y)\right].
\label{P42}
\end{eqnarray}
Note that for $1 \leq q < 2$ the first term on the r.h.s of Eqs. (\ref{P32}) and (\ref{P42}) is absent. For these values of $q$ the Wightman function for the massless scalar field above is in agreement, for $D=3$, with the propagator obtained in Ref. \cite{DeLorenci:2002jv}\footnote{Note that the authors in Ref. \cite{DeLorenci:2002jv} calculated the Feynman propagator. Thus, our result in (\ref{P42}) coincides with theirs as long as we use the relation $W(x,x')=iG^{(\alpha,\kappa)}(x,x')$, where $G^{(\alpha,\kappa)}(x,x')$ is the Feynman propagator in their notation.}. However, here we have provided closed expressions for all values of $q$, Eqs. (\ref{P32}) and (\ref{P42}), for massive and massless scalar fields in a high-dimensional cosmic dispiration spacetime, respectively.

%
\section{Vacuum expectation value of the field squared}
\label{sec3}
The VEV of the field squared, $\left\langle \Phi^2\right\rangle$, is formally obtained from the Wightman functions (\ref{P32}) in the coincidence limit $x'\rightarrow x$, i.e,
\begin{equation}
\left\langle \Phi^2\right\rangle =\lim_{x'\rightarrow x}W(x,x').
 \label{VEV}
\end{equation}
However, the Wightman function (\ref{P32}) is divergent in this limit. Its divergent contribution is given by the Hadamard function obtained from the $l=0$ term and, according to the standard prescription discussed in Refs. \cite{Christensen:1976vb, Wald:1978pj}, it must be subtracted from the Wightman function $W(x,x')$. So, the renormalized result of the above VEV is
\begin{equation}
\left\langle \Phi^2\right\rangle_{{\rm ren}}=\lim_{x'\rightarrow x}\left[W(x,x')-G_{\rm H}(x,x')\right],
 \label{renVEV}
\end{equation}
where the Hadamard function $G_{\rm H}(x,x')$ is given by
\begin{equation}
G_{\rm H}(x,x')=\frac{m^{(D-1)}}{(2\pi)^{\frac{(D+1)}{2}}}f_{\frac{D-1}{2}}\left(m\sigma_0\right),
\label{Had1}
\end{equation}
with $\sigma_0$ defined as in (\ref{GD2}) for $l=0$, i.e, 
\begin{eqnarray}
\sigma_0^2 = -\Delta t^2 + \Delta {\bf r}_{\parallel}^2 + r'^2 + r^2 - 2rr'\cos\left(\Delta\varphi\right) + \Delta Z^2.
\label{GD0}
\end{eqnarray}
Here we would like to call attention to the fact that the Hadamard function above is the solution of the corresponding homogeneous differential equation in the coordinate system defined by (\ref{Me2}), when $q=1$.

The renormalized VEV of the field squared (\ref{renVEV}) is, therefore, found to be 
\begin{eqnarray}
\left\langle \Phi^2\right\rangle_{{\rm ren}}&=&\frac{2m^{(D-1)}}{(2\pi)^{\frac{(D+1)}{2}}}\left[\sideset{}{'}\sum_{l=1}^{[q/2]} f_{\frac{D-1}{2}}\left(m\sqrt{(\bar{\kappa} l)^2+4r^2s_l^2}\right)\right.\nonumber\\
&-&\left.\frac{q}{\pi^2}\sum_{n=-\infty}^{\infty}\int_{0}^{\infty}dyf_{\frac{D-1}{2}}\left(m\sqrt{(\bar{\kappa} n)^2+4r^2\cosh^2(y)}\right)M_{n,q}(2y)\right],
 \label{renVEV2}
\end{eqnarray}
where $M_{n,q}(0,2y) = M_{n,q}(2y)$ and $[q/2]$ represents the integer part of $q/2$, and the prime on the sign of the summation means that in the case $q/2$ is integer the term $l = q/2$ should be taken with the coefficient 1/2. Note that we have also made the change, $y\rightarrow 2y$, and defined 
\begin{eqnarray}
s_l=\sin(\pi l/q).
\label{sl}
\end{eqnarray}
Some limiting cases of Eq. (\ref{renVEV2}) are possible to be considered for $D=3$. For instance, for fixed $m\bar{\kappa}$, in the limit $mr\gg 1$, we can ignore the summation indices $l$ and $n$ in the argument of the $f_{\mu}(x)$ functions. This procedure can only be adopted on the second term on the r.h.s of Eq.  (\ref{renVEV2}) because the function $M_{n,q}(2y)$ presents a negligible contribution for large values of $n$. Thereby, this provides, as a leading contribution, the expression of the renormalized VEV of the field squared in the pure cosmic string spacetime. In other words, far away from the cosmic dispiration the screw dislocation is negligible. On the other hand, keeping $mr$ fixed, the leading contribution in the limit $m\bar{\kappa}\gg 1$ comes from $n=0$ in the second term on the r.h.s of (\ref{renVEV2}). For the first term, we can neglect the term with $s_l$ inside the brackets. Thus, in this regime we obtain the following approximation:
\begin{eqnarray}
\left\langle \Phi^2\right\rangle_{{\rm ren}}&\simeq&\frac{m^2}{2\pi^2}\left[\sqrt{\frac{\pi}{2}}\frac{e^{-m\bar{\kappa}}}{(m\bar{\kappa})^{\frac{3}{2}}} - \left.\frac{1}{mr}\int_{0}^{\infty}dy\frac{K_1\left(2mr\cosh y\right)}{\cosh y}\frac{1}{(\pi^2+4y^2)}\right]\right.\nonumber\\
&\simeq&\left.\frac{m}{2\pi^2 r}\int_{0}^{\infty}dy\frac{K_1\left(2mr\cosh y\right)}{\cosh y}\frac{1}{(\pi^2+4y^2)}\right].
 \label{appVEV2}
\end{eqnarray}
Note that, in this approximation, the integral term is the dominant one since the first term is exponentially suppressed. As a consequence, the final result for Eq. (\ref{appVEV2}) is independent of $\kappa$.

Additionally, in the limit $m\rightarrow 0$, Eq. (\ref{renVEV2}) leads to the expression of the renormalized VEV of the massless scalar field squared, i.e,
\begin{eqnarray}
\left\langle \Phi^2\right\rangle_{{\rm ren}}&=&\frac{\Gamma\left(\frac{D-1}{2}\right)}{2\pi^{\frac{(D+1)}{2}}\bar{\kappa}^{D-1}}\left[\sideset{}{'}\sum_{l=1}^{[q/2]} \left[l^2+\epsilon^2 s_l^2\right]^{\frac{1-D}{2}}\right.\nonumber\\
&-&\left.\frac{q}{\pi^2}\sum_{n=-\infty}^{\infty}\int_{0}^{\infty}dy \left[n^2+\epsilon^2\cosh^2(y)\right]^{\frac{1-D}{2}}M_{n,q}(2y)\right],
\label{renFS}
\end{eqnarray}
where $\epsilon=2r/\bar{\kappa}=qr/\pi\kappa$. The result in (\ref{renFS}) for the VEV of the field squared in the massless case, coincides, for $D=3$, with the one obtained in Ref. \cite{DeLorenci:2002jv} when $q<2$. However we would like to emphasize that we have obtained, for all possible values of $q$, closed general expressions for the Wightman function (\ref{P32}) and, consequently, for the the renormalized VEV of the field squared (\ref{renVEV}) for the massive case, in a high-dimensional cosmic dispiration spacetime. We have also formally demonstrated in the Appendix \ref{appB} that when $\kappa=0$ we recover the Wightman function in the pure cosmic string spacetime and all the resulting expressions. We would like to point out that the general result above has never been obtained before. In \cite{DeLorenci:2002jv}, the VEV of the field squared was obtained for a massless case and only for $q<2$.

For the three-dimensional case, $D=3$, the summation in $n$ present in Eq. (\ref{renFS}) provides a closed expression and we find that the renormalized VEV of the massless field squared is written as
\begin{eqnarray}
\left\langle \Phi^2\right\rangle_{{\rm ren}}=\frac{1}{2\pi^2\bar{\kappa}^{2}}\left[\sideset{}{'}\sum_{l=1}^{[q/2]} \left[l^2+\epsilon^2 s_l^2\right]^{-1}
-\frac{q}{\pi^2}\int_{0}^{\infty}g(y,\epsilon,q)dy \right],
\label{renFS2}
\end{eqnarray}
where
\begin{eqnarray}
g(y,\epsilon,q)&=&\frac{2\pi^3}{\epsilon f(y,q)\cosh(y)}\left\{q \left[4 \epsilon^2\pi^2 \cosh^2(y) + q^2 (\pi^2 + 4 y^2)\right]\coth\left(\epsilon\pi\cosh(y)\right)\right. \nonumber\\
&+&\left. \frac{2\epsilon\cosh(y)\left[4 \epsilon^2\pi^2 \cosh^2(y) + q^2 (\pi^2 - 4 y^2)\right] \sin(q\pi) - 
 8q^2 \epsilon \pi y \cosh(y) \sinh(2 q y)}{\cosh(2 q y)-\cos(\pi q)}\right\},\nonumber\\
\label{func1}
\end{eqnarray}
with
\begin{eqnarray}
f(y,q)= 16 \epsilon^4\pi^4 \cosh^4(y)  + 8q^2  \epsilon^2\pi^2 (\pi^2 - 4 y^2) \cosh^2(y)  +  q^4 (\pi^2 + 4 y^2)^2.
\label{fun2}
\end{eqnarray}

Let us now investigate the limiting cases $\epsilon\ll 1$ and $\epsilon\gg 1$ of Eq. (\ref{renFS2}).  Thereby, in the limit $\epsilon\gg 1$, we recover the result for the VEV of the field squared, in the massless case, in the pure cosmic string spacetime \cite{BezerradeMello:2011sm,BezerradeMello:2011nv} using the expression
\begin{eqnarray}
\lim_{\epsilon\rightarrow\infty}\epsilon^2g(y,\epsilon,q)=\frac{\pi}{\cosh^2(y)}\frac{\sin(q\pi)}{\left[\cosh(2 q y)-\cos(\pi q)\right]}.
\label{lim}
\end{eqnarray}
In other words, the screw dislocation effects are negligible in the regime where $r\gg\pi\kappa/q$.

On the other hand, in the opposite limit, when $\epsilon\ll 1$, one can expand Eq. (\ref{func1}) in power series for small $\epsilon$. In this case, the main contribution for the renormalized VEV of the massless field squared, Eq. (\ref{renFS2}), is found to be
\begin{eqnarray}
\left\langle \Phi^2\right\rangle_{{\rm ren}}\simeq -\frac{1}{48\pi^2r^2}g_0 + O(\epsilon^2)
\label{app1}
\end{eqnarray}
where 
\begin{eqnarray}
g_0 \simeq 12\int_{0}^{\infty}dy \frac{1}{\left(\pi^2+4y^2\right)}\frac{1}{\cosh^2(y)}\simeq 1,
\label{num1}
\end{eqnarray}
 is obtained numerically.

We would like now to call attention to a very interesting fact that has already been pointed out by the authors in Ref. \cite{DeLorenci:2002jv}. For this purpose, let us consider $q<2$ and $\kappa=0$, where in this case the first term on the r.h.s of (\ref{renFS}) is absent. The VEV of the field squared in the cosmic string spacetime is then given by 
\begin{eqnarray}
\left\langle \Phi^2\right\rangle_{{\rm ren}}=-\frac{q\sin(q\pi)}{8\pi^3r^2}\int_{0}^{\infty}dy \frac{1}{\left[\cosh(2yq)-\cos(q\pi)\right]}\frac{1}{\cosh^2(y)}.
\label{renCS}
\end{eqnarray}
By noting that the main contribution for the expression above comes from small values of $y$, we can obtain an approximation for $q\ll 1$ as 
\begin{eqnarray}
\left\langle \Phi^2\right\rangle_{{\rm ren}}&\simeq&-\frac{1}{4\pi^2r^2}\int_{0}^{\infty}dy \frac{1}{\left(\pi^2+4y^2\right)}\frac{1}{\cosh^2(y)}\nonumber\\
&\simeq& -\frac{1}{48\pi^2r^2}g_0.
\label{renCS2}
\end{eqnarray}
Thereby, we can see that the approximation for the expression (\ref{renCS}) when $q\ll 1$ gives, as the main contribution, the expression in  (\ref{renCS2}). The latter is equivalent to the leading contribution in Eq.  (\ref{app1}) obtained when $\kappa\neq 0$ and $\epsilon\ll 1$. This can also be verified through the alternative analytic expression, given by Eq. (14) in Ref. \cite{DeLorenci:2002jv}, for the VEV of the field squared in the cosmic string spacetime. 

%
\section{Energy-momentum tensor}
\label{sec4}
%
Since we have obtained the Wightman function (\ref{P32}) and the the renormalized VEV of the field squared, Eq. (\ref{renVEV2}), we are now in position to calculate the VEV of the energy-momentum tensor in the high-dimensional cosmic dispiration spacetime. By adopting the coordinate system in which the metric (\ref{Me2}) is written, the formal expression for the components of the energy-momentum tensor can be calculated by using the formula \cite{BezerradeMello:2011nv}
\begin{eqnarray}
\left\langle T_{\mu\nu}\right\rangle = \lim_{x'\rightarrow x}\partial_{\mu'}\partial_{\nu}W(x,x') + \left[\left(\xi - 1/4\right)g_{\mu\nu}\Box - \xi\nabla_{\mu}\nabla_{\nu} - \xi R_{\mu\nu}\right]\left\langle \Phi^2\right\rangle.
\label{EMT}
\end{eqnarray}
However, in this spacetime, the Ricci tensor, $R_{\mu\nu}$, presents a Dirac delta-like distribution at origin. Thereby, in order to avoid singular behaviours in the calculation of the VEV of the energy-momentum tensor, and also to be compatible with the calculations performed in the previous sections, we will adopt in this section $\xi = 0$. Thus, using Eqs. (\ref{P32}) and (\ref{renVEV2}) it is possible to calculate the renormalized VEV of the energy-momentum tensor for a minimally coupled scalar field, as we shall see below. 

Let us then start by calculating, $\Box\left\langle \Phi^2\right\rangle_{{\rm ren}}$. In this case the only derivative contribution in the d'Alembertian operator is the one due to the radial component, $r$. So, we have
\begin{eqnarray}
\Box\left\langle \Phi^2\right\rangle_{{\rm ren}} &=& -\frac{8m^{(D+1)}}{(2\pi)^{\frac{(D+1)}{2}}}\left\{\sideset{}{'}\sum_{l=1}^{[q/2]} \left[(2mr)^2s_l^4f_{\frac{D+3}{2}}(\lambda_l)-2s_l^2f_{\frac{D+1}{2}}(\lambda_l)\right]\right.\nonumber\\
&-&\left.\frac{q}{\pi^2}\sum_{n=-\infty}^{\infty}\int_{0}^{\infty}dy\left[(2mr)^2\cosh^4(y)f_{\frac{D+3}{2}}(\lambda_{n})-2\cosh^2(y)f_{\frac{D+1}{2}}(\lambda_{n})\right]M_{n,q}(2y)\right\}.\nonumber\\
\label{deri}
\end{eqnarray}
%
%
\begin{figure}[!htb]
\begin{center}
\includegraphics[width=0.4\textwidth]{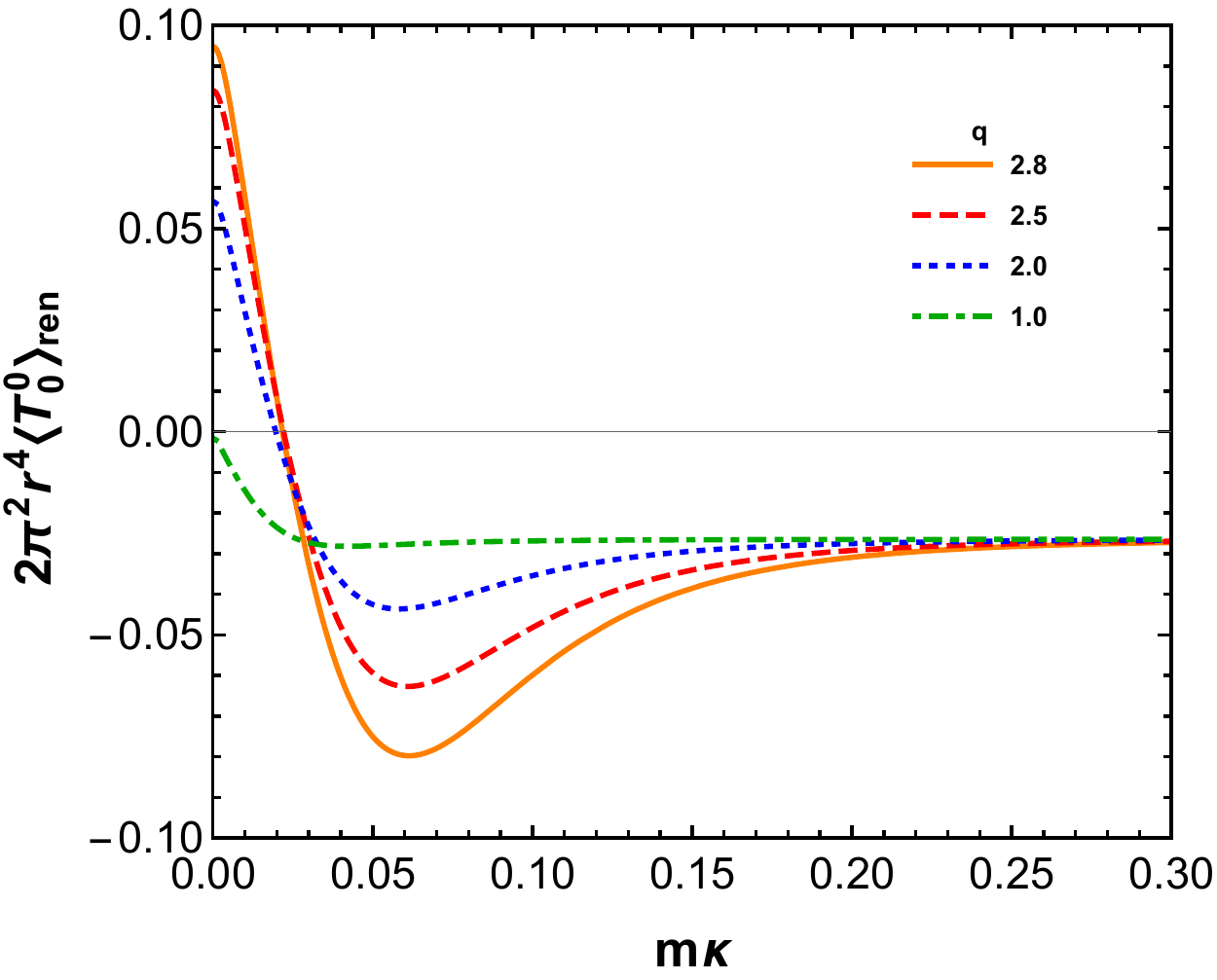}
%
\includegraphics[width=0.41\textwidth]{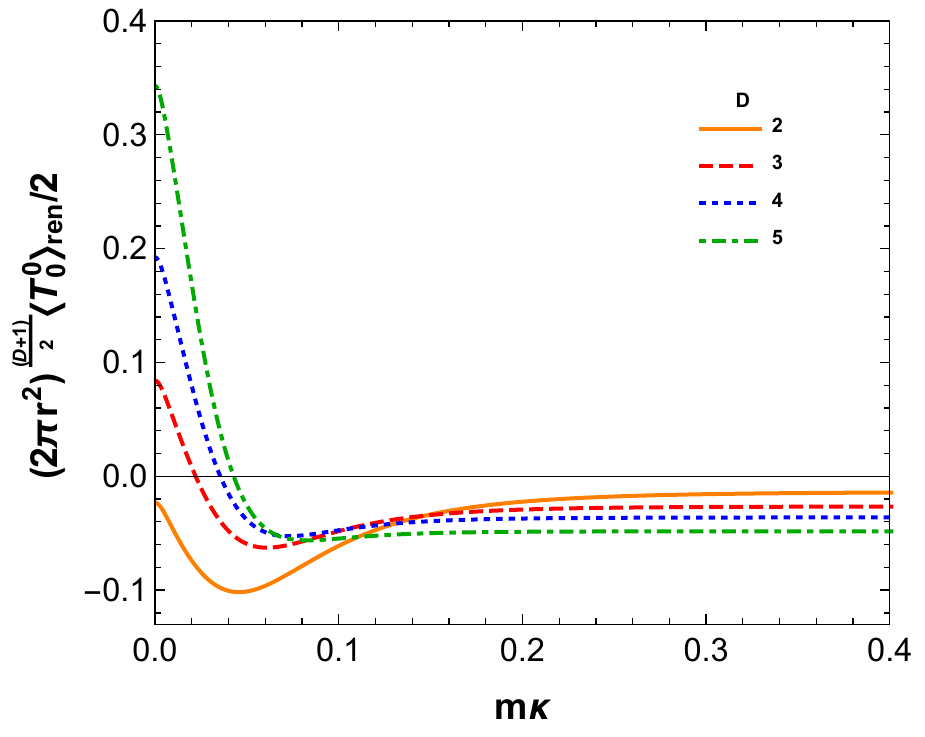}
\caption{\small{On the left: for $D=3$, the (0,0)-component of the energy-momentum tensor, $\left\langle T^{0}_{0}\right\rangle_{{\rm ren}}$, multiplied by $2\pi^2 r^4$, in the massive scalar field case, is plotted as a function of $m\kappa$ for different values of the cosmic string parameter, $q$, considering $mr=0.1$. On the right: the energy-momentum tensor, $\left\langle T^{0}_{0}\right\rangle_{{\rm ren}}$, multiplied by $\frac{(2\pi r^2)^{\frac{(D+1)}{2}}}{2}$, in the massive scalar field case, is plotted as a function of $mk$ for different values of the spatial dimension $D$, considering $q=2.5$ and $mr = 0.1$. Both plots are for $\xi=0$.}}
\label{f1}
\end{center}
\end{figure}
%
Moreover, it is also necessary to calculate the ordinary derivative of $W(x,x')$ with respect to the spacetime coordinates. Thus, taking the Wightman function (\ref{P32}), we obtain after long and intermediate steps the renormalized VEV of the energy-momentum tensor for the massive scalar field
%
%
\begin{figure}[!htb]
\begin{center}
\includegraphics[width=0.4\textwidth]{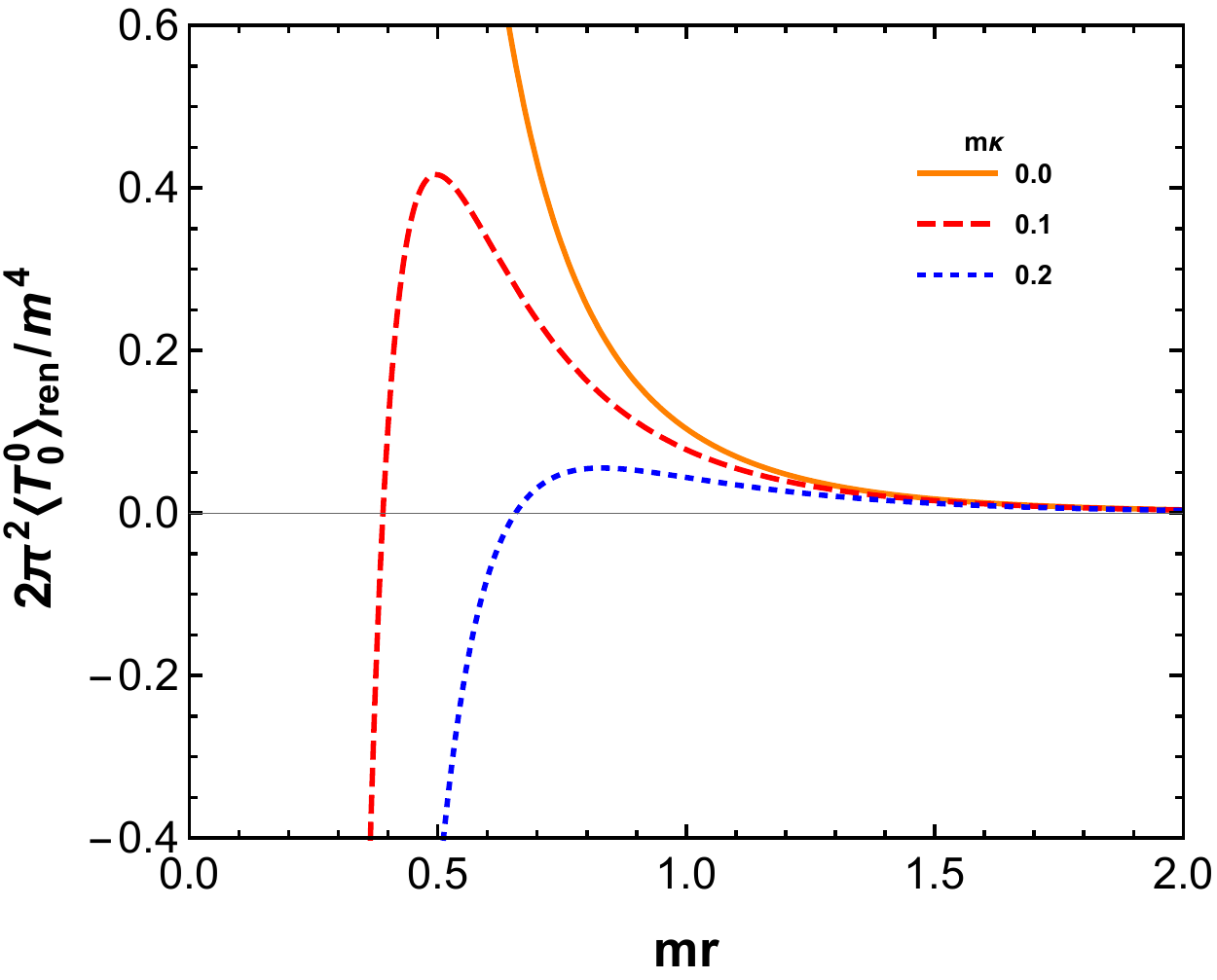}
%
\includegraphics[width=0.408\textwidth]{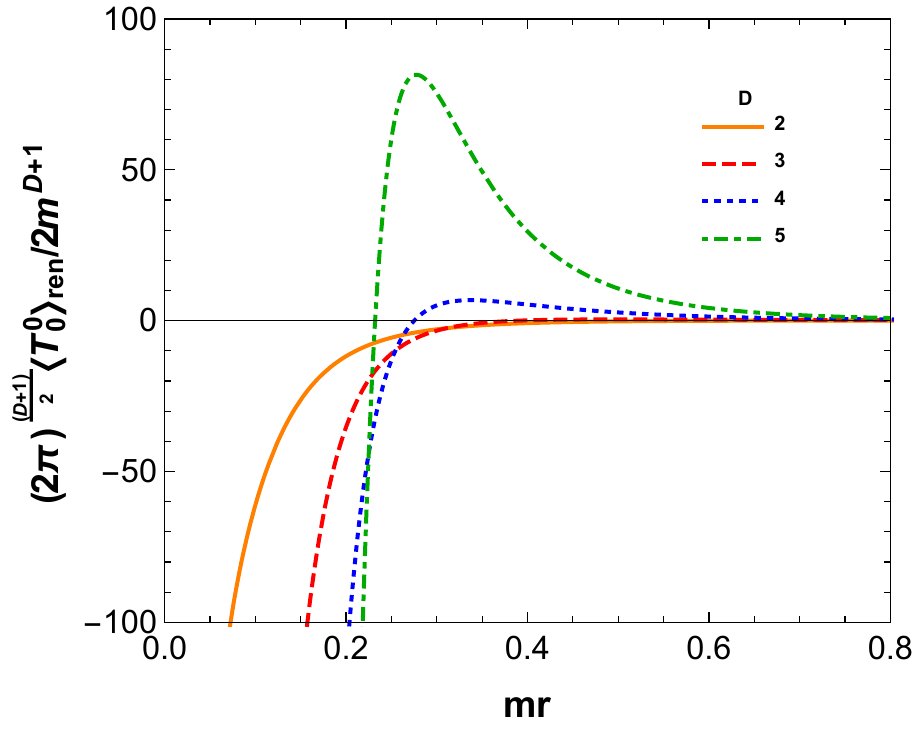}
\caption{\small{On the left: for $D=3$, the (0,0)-component of the energy-momentum tensor, $\left\langle T^{0}_{0}\right\rangle_{{\rm ren}}$, in units of $m^4/2\pi^2$, in the massive scalar field case, is plotted as a function of $mr$ for different values of $m\kappa$, considering $q=2.5$. On the right: the energy-momentum tensor, $\left\langle T^{0}_{0}\right\rangle_{{\rm ren}}$, multiplied by $\frac{(2\pi)^{\frac{(D+1)}{2}}}{2m^{D+1}}$, in the massive scalar field case, is plotted as a function of $mr$ for different values of the spatial dimension $D$, considering $q=2.5$ and $m\kappa = 0.1$. Both plots are for $\xi=0$.}}
\label{f2}
\end{center}
\end{figure}
%
\begin{eqnarray}
\left\langle T^{\mu}_{\nu}\right\rangle_{{\rm ren}}  &=&\frac{2m^{(D+1)}}{(2\pi)^{\frac{(D+1)}{2}}}\left[\sideset{}{'}\sum_{l=1}^{[q/2]}F^{\mu}_{\nu,l}(2mr,m\bar{\kappa},s_l)\right.\nonumber\\
&-&\left.\frac{q}{\pi^2}\sum_{n=-\infty}^{\infty}\int_{0}^{\infty}dyF^{\mu}_{\nu,n}(2mr,m\bar{\kappa},\cosh y)M_{n,q}(2y)\right],
\label{EMT2}
\end{eqnarray}
where, for the diagonal components, the functions $F^{\mu}_{\nu,\sigma}(u,g,v)$ are defined as
\begin{eqnarray}
F_{0,\sigma}^0(u,g,v) &=& u^2v^4f_{\frac{D+3}{2}}(\lambda_{\sigma}) - \left[2v^2 + 1\right]f_{\frac{D+1}{2}}(\lambda_{\sigma}),\nonumber\\
F_{1,\sigma}^1(u,g,v) &=& - f_{\frac{D+1}{2}}(\lambda_{\sigma}),\nonumber\\
F_{2,\sigma}^2(u,g,v) &=& u^2v^2f_{\frac{D+3}{2}}(\lambda_{\sigma}) - f_{\frac{D+1}{2}}(\lambda_{\sigma}),\nonumber\\
F_{3,\sigma}^3(u,g,v) &=&F_{0,\sigma}^0(u,g,v) + (\bar{\kappa} m\sigma)^2f_{\frac{D+3}{2}}(\lambda_{\sigma}),\nonumber\\
F_{i,\sigma}^{i}(u,g,v)&=&F_{0,\sigma}^0(u,g,v),\qquad \mathrm{for}\;\;\;\; i=4,...,D.
\label{tt}
\end{eqnarray}
As to the off-diagonal component, its expression is given by
\begin{eqnarray}
\left\langle T^{3}_{2}\right\rangle_{{\rm ren}}  &=&\frac{2m^{(D+3)}}{(2\pi)^{\frac{(D+1)}{2}}}\bar{\kappa}r^2\left[\sideset{}{'}\sum_{l=1}^{[q/2]}l\sin(2\pi l/q)f_{\frac{D+3}{2}}(\lambda_{l})\right.\nonumber\\
&-&\left.\frac{q^2}{\pi^3}\sum_{n=-\infty}^{\infty}\int_{0}^{\infty}dyy\sinh(2y)p(y,n,q)f_{\frac{D+3}{2}}(\lambda_{n})\right],
\label{EMT2cr}
\end{eqnarray}
with
\begin{eqnarray}
p(y,n,q)=\frac{n}{\left[\left(n+\frac{q}{2}\right)^2+\left(\frac{qy}{\pi}\right)^2\right]},
\label{funp}
\end{eqnarray}
and the argument $\lambda_{\sigma}$ in Eqs.  (\ref{tt}) and (\ref{EMT2cr}) is written as
\begin{eqnarray}
\lambda_{\sigma}=\sqrt{u^2v^2 + (g\sigma)^2},
\label{arg}
\end{eqnarray}
%
\begin{figure}[!htb]
\begin{center}
\includegraphics[width=0.4\textwidth]{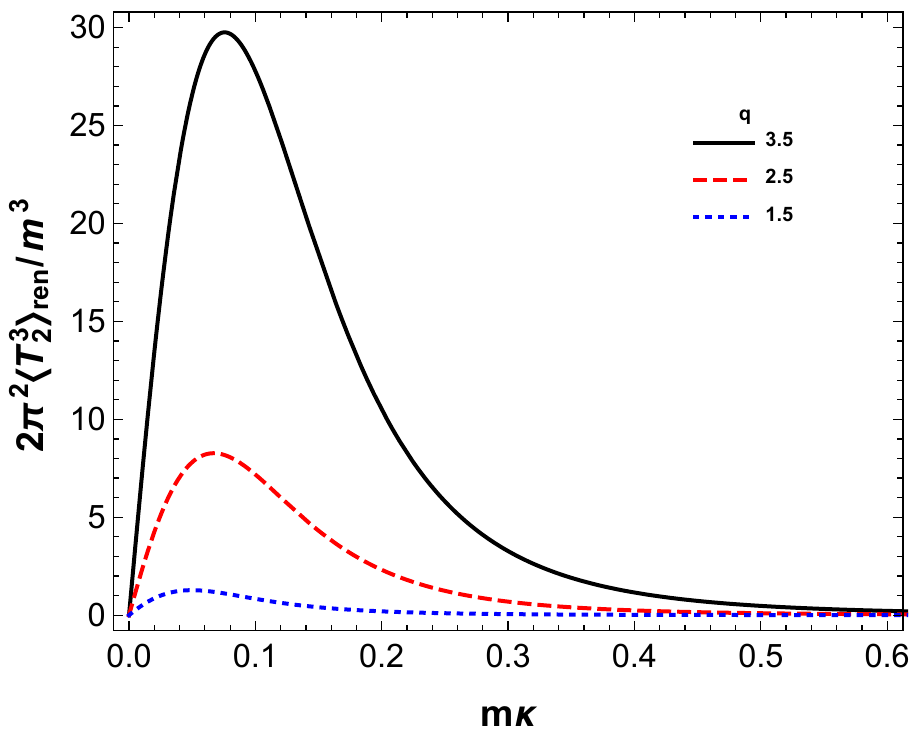}
%
\includegraphics[width=0.4055\textwidth]{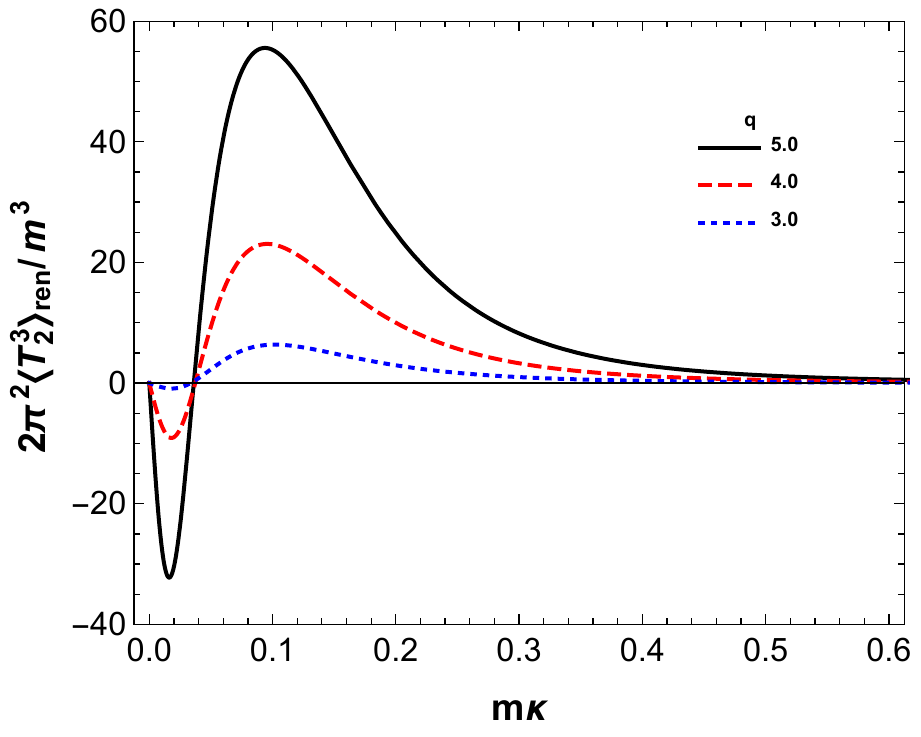}
\caption{\small{For $D=3$, the off-diagonal component of the energy-momentum tensor, $\left\langle T^{3}_{2}\right\rangle_{{\rm ren}}$, in units of $m^3/2\pi^2$, in the massive scalar field case, is plotted as a function of $m\kappa$ for different values of $q$, considering $mr = 0.2$.}}
\label{f3}
\end{center}
\end{figure}
%
where the greek letter $\sigma$ stands for $l$ in the first term in (\ref{EMT2}) and for $n$ in the second term. The boost invariance in the directions $x^{i}$, $i=4, ... ,D$ provides us with (no summation over $i$) $F_{i,\sigma}^{i}(u,g,v)=F_{0,\sigma}^0(u,g,v)$. On the other hand, because the boost invariance in the $z$-direction is broken by the presence of the torsion, we have that  $F^3_{3,\sigma}(u,g,v)\neq F_{0,\sigma}^0(u,g,v)$, as we can see from Eq. (\ref{tt}). Note that in order to calculate the $(2,2)$-component of the energy-momentum tensor we used Eq. (52) from Ref. \cite{BezerradeMello:2011sm}. Moreover, we have obtained the expression (\ref{EMT2cr}) for the off-diagonal component by deriving the Wightman function (\ref{P32}) with respect to $Z'$ and $\varphi$, and integrating by parts in $y$ afterwards. The expressions \eqref{EMT2}-\eqref{arg} for the energy-momentum tensor, to the best of our knowledge, are new results. Again, we would like to emphasize that the authors in Ref. \cite{DeLorenci:2002jv} only considered the energy-momentum tensor associated with a massless scalar field in $(3+1)$-dimensional cosmic dispiration spacetime, for values $q<2$. In fact, the authors do not not even provide closed expressions for this case but, instead, an approximated expression for larges values of $r$. In contrast, the expressions \eqref{EMT2}-\eqref{arg} we have presented are exact closed ones for the massive scalar field case, valid for all possible values of the cosmic string parameter $q$. Besides, as we will see below, one can obtain an exact closed expression for the energy-momentum tensor in the massless case by taking the limit $m\rightarrow 0$.

In Fig.\ref{f1} we have plotted on the left, for $D=3$, the $(0,0)$-component of the energy-momentum tensor multiplied by  $2\pi^2 r^4$, with respect to $m\kappa$ for different values of $q$ and taking $mr=0.1$. The plot on the right shows the energy-momentum tensor, multiplied by $\frac{(2\pi r^2)^{\frac{(D+1)}{2}}}{2}$, in the massive scalar field case, in terms of $mk$ for different values of the spatial dimension $D$, considering $q=2.5$ and $mr = 0.1$. In Fig.\ref{f2}, on the left, for $D=3$, one can see the plot of $\left\langle T^{0}_{0}\right\rangle_{{\rm ren}}$, in units of $m^4/2\pi^2$, with respect to $mr$ for different values of $m\kappa$ and taking $q=2.5$ whilst the plot on the right shows the energy-momentum tensor, multiplied by $\frac{(2\pi)^{\frac{(D+1)}{2}}}{2m^{D+1}}$, in the massive scalar field case, in terms of $mr$ for different values of the spatial dimension $D$, considering $q=2.5$ and $m\kappa = 0.1$. It is clear that, compared to $m\kappa =0$, the presence of the torsion changes the behaviour of $\left\langle T^{0}_{0}\right\rangle_{{\rm ren}}$ near the string significantly. Furthermore, for $D=3$, the plot of the off-diagonal component $\left\langle T^{3}_{2}\right\rangle_{{\rm ren}}$, in units of $m^3/2\pi^2$, with respect to $m\kappa$ is shown in Fig.\ref{f3}. The curves are obtained for different values of $q$ and taking $mr =0.2$. Note that Figs.\ref{f1} and \ref{f2} shows the plots for the minimally coupled scalar field.

We have proved that the energy-momentum tensor (\ref{EMT2}) satisfies two important requirements: the covariant conservation condition $\nabla_{\mu}\left\langle T^{\mu}_{\nu}\right\rangle_{{\rm ren}} = 0$, which can be reduced down to the relation 
\begin{equation}
\left\langle T^{2}_{2}\right\rangle_{{\rm ren}} =\partial_r\left(r\left\langle T^{1}_{1}\right\rangle_{{\rm ren}}\right), 
\label{rel1}
\end{equation}
as well as the trace identity 
\begin{eqnarray}
\left\langle T^{\mu}_{\mu}\right\rangle_{{\rm ren} }=D(\xi - \xi_D)\nabla_{\mu}\nabla^{\mu}\left\langle \Phi^2\right\rangle_{{\rm ren} } + m^2\left\langle \Phi^2\right\rangle_{{\rm ren} }.
\label{trace}
\end{eqnarray}
Note that, in our case, considering the components of the energy-momentum tensor, \eqref{EMT2}-\eqref{tt}, we have proved that the identity \eqref{trace} is satisfied for $\xi = 0$. Now we want to consider the asymptotic behaviours of both the $(0,0)$ and the off-diagonal components of Eq. (\ref{EMT2}), for $D=3$, in the regimes $m\bar{\kappa}\gg mr$ and $mr\gg m\bar{\kappa}$. In the latter, based on the same argument we used for the VEV of the field squared, (\ref{renVEV2}), the leading contribution to the $(0,0)$-component of the energy-momentum tensor coincides with the corresponding one in the cosmic string spacetime \cite{BezerradeMello:2011nv, BezerradeMello:2011sm}.  As to the regime where $m\kappa\gg mr$, also assuming $mr\ll 1$, we find
\begin{eqnarray}
\left\langle T^{0}_{0}\right\rangle_{{\rm ren}} &\simeq&\frac{m^{4}}{2\pi^{2}}\left[\sideset{}{'}\sum_{l=1}^{[q/2]}F^{0}_{0,l}(m\bar{\kappa}l)-\frac{1}{4m^4r^4}\int_{0}^{\infty}dy\frac{\left[2\cosh^2(y) - 1\right]}{\cosh^4(y)(\pi^2 + 4y^2)}\right]\nonumber\\
&\simeq&\frac{m^{4}}{2\pi^{2}}\sideset{}{'}\sum_{l=1}^{[q/2]}F^{0}_{0,l}(m\bar{\kappa}l)-\frac{1}{75\pi^2r^4},
\label{asym1}
\end{eqnarray}
where $F^{0}_{0,l}(2mr,m\bar{\kappa}, s_l)=F^{0}_{0,l}(m\bar{\kappa})$ is defined as in (\ref{tt}) but with $\lambda_l\simeq m\bar{\kappa}l$. Note that we have used a small argument approximation for the Macdonald function $K_{\mu}(x)$ \cite{Abramowitz}. Note also that the contribution on the second term on the r.h.s of Eq. (\ref{asym1}) comes from the $n=0$ term in the sum of Eq.  (\ref{EMT2}). Moreover, the additional requirement, $m\bar{\kappa}\gg 1$, along with a large argument approximation for $K_{\mu}(x)$ \cite{Abramowitz}, makes the first term on the r.h.s of Eq. (\ref{asym1}) to be exponentially suppressed by the factor $\frac{e^{-m\bar{\kappa}}}{(m\bar{\kappa})^{\frac{5}{2}}}$ and, as a consequence, the second term is the dominant one. This asymptotic behaviour is shown in Figs.\ref{f1},\ref{f2}.

On the other hand, for the off-diagonal component, in the regime $mr\gg m\bar{\kappa}$, with $mr\gg 1$, we obtain  
\begin{eqnarray}
\left\langle T^{3}_{2}\right\rangle_{{\rm ren}} &\simeq&\frac{m^4\kappa}{4\pi q(2mr)^{\frac{3}{2}}}\sqrt{\frac{\pi}{2}}\left[\sideset{}{'}\sum_{l=1}^{[q/2]}\frac{l\sin(2l\pi/q)}{\sin^{\frac{7}{2}}(l\pi/q)}e^{-2mr\sin(l\pi/q)}\right.\nonumber\\
&-&\left.\frac{q^2}{\pi^3}\int_{0}^{\infty}dy\frac{y\sinh(2y)}{\cosh^{\frac{7}{2}}(y)}b(y,q)e^{-2mr\cosh y}\right],
\label{asym2}
\end{eqnarray}
where 
\begin{eqnarray}
b(y,q)&=&\sum_{n=-\infty}^{\infty}p(y,n,q)\nonumber\\
&=& \frac{1}{2y}\frac{\left(2\pi y\sin(q\pi)-\pi^2\sinh(2qy)\right)}{\left[\cosh(2qy)-\cos(q\pi)\right]}.
\label{asym3}
\end{eqnarray}
The asymptotic behaviour described by Eq. (\ref{asym2}) can be seen in Fig.\ref{f3}.

Furthermore, in the regime where $m\bar{\kappa}\gg mr$, also with $m\bar{\kappa}\gg 1$, the off-diagonal component is given by 
\begin{eqnarray}
\left\langle T^{3}_{2}\right\rangle_{{\rm ren}}  &=&\frac{m^3}{2\pi^2}(mr)^2\sqrt{\frac{\pi}{2}}\left[\sideset{}{'}\sum_{l=1}^{[q/2]}\frac{\sin(2\pi l/q)e^{-m\bar{\kappa}l}}{(m\bar{\kappa}l)^{\frac{5}{2}}}-\frac{m\bar{\kappa}q^2}{\pi^3}\sum_{n=-\infty}^{\infty}e^{-m\bar{\kappa}n}\right.\nonumber\\
&\times&\left.\int_{0}^{\infty}dy\frac{y\sinh(2y)p(y,n,q)}{[(2mr)^2\cosh^2(y)+(m\bar{\kappa}n)^{2}]^{\frac{7}{2}}}\right].
\label{AEMT3.1}
\end{eqnarray}
In this regime, for the values of $q>2$, the first term is the dominant one and $\left\langle T^{3}_{2}\right\rangle_{{\rm ren}}$ goes to zero from above while for $q<2$ the first term is absent and $\left\langle T^{3}_{2}\right\rangle_{{\rm ren}}$ also goes to zero from above.
This behavior is explicitly exhibited in Fig.\ref{f3}. One should note that, in order to get the approximation \eqref{AEMT3.1}, we made use of a large argument approximation for the Macdonald function, also assuming that $e^{-\sqrt{(2mr)^2\cosh^2(y)+(m\bar{\kappa}n)^{2}}} < e^{-m\bar{\kappa}n}$, for given values of the argument.

For the massless scalar field case, the VEV of the energy-momentum tensor can be obtained from Eqs. (\ref{EMT2}) and (\ref{tt})  in the limit $m\rightarrow 0$. This provides 
\begin{eqnarray}
\left\langle T^{\mu}_{\nu}\right\rangle_{{\rm ren}}  &=&\frac{\Gamma\left(\frac{D+1}{2}\right)\epsilon^{(D+1)}}{(2r)^{D+1}\pi^{\frac{(D+1)}{2}}}\left[\sideset{}{'}\sum_{l=1}^{[q/2]}\frac{H^{\mu}_{\nu,l}(\epsilon,s_l)}{\left[\epsilon^2s_l^2+l^2\right]^{\frac{D+3}{2}}}\right.\nonumber\\
&-&\left.\frac{q}{\pi^2}\sum_{n=-\infty}^{\infty}\int_{0}^{\infty}dy\frac{H^{\mu}_{\nu,n}(\epsilon,\cosh y)}{\left[\epsilon^2\cosh^2(y)+n^2\right]^{\frac{D+3}{2}}}M_{n,q}(2y)\right],
\label{EMT3}
\end{eqnarray}
where, for the diagonal components, the new  functions $H_{\nu,\sigma}^{\mu}(\epsilon,v)$ are defined as

\begin{eqnarray}
H_{0,\sigma}^{0}(\epsilon,v) &=& \epsilon^2v^4(D+1)- \left[2v^2 + 1\right](\epsilon^2v^2+\sigma^2),\nonumber\\
H_{1,\sigma}^{1}(\epsilon,v) &=& - (\epsilon^2v^2+\sigma^2),\nonumber\\
H_{2,\sigma}^{2}(\epsilon,v) &=& (\epsilon^2v^2D-\sigma^2),\nonumber\\
H_{3,\sigma}^{3}(\epsilon,v) &=&H_{0,\sigma}^{0}(u,v) + \sigma^2(D+1),\nonumber\\
H_{i,\sigma}^{i}(\epsilon,v)&=&H_{0,\sigma}^0(\epsilon,v),\qquad \mathrm{for}\;\;\;\; i=4,...,D.
\label{tt2}
\end{eqnarray}
The off-diagonal component in this case is given by
\begin{eqnarray}
\left\langle T^{3}_{2}\right\rangle_{{\rm ren}}  &=&\frac{\Gamma\left(\frac{D+1}{2}\right)\epsilon^{(D+3)}}{(2r)^{D+1}\pi^{\frac{(D+1)}{2}}}\frac{(D+1)\bar{\kappa}}{4}\left[\sideset{}{'}\sum_{l=1}^{[q/2]}\frac{l\sin(2\pi l/q)}{\left[\epsilon^2s_l^2+l^2\right]^{\frac{D+3}{2}}}\right.\nonumber\\
&-&\left.\frac{q^2}{\pi^3}\sum_{n=-\infty}^{\infty}\int_{0}^{\infty}dy\frac{y\sinh(2y)p(y,n,q)}{\left[\epsilon^2\cosh^2(y)+n^2\right]^{\frac{D+3}{2}}}\right],
\label{EMT3cr}
\end{eqnarray}
%
\begin{figure}[!htb]
\begin{center}
\includegraphics[width=0.4\textwidth]{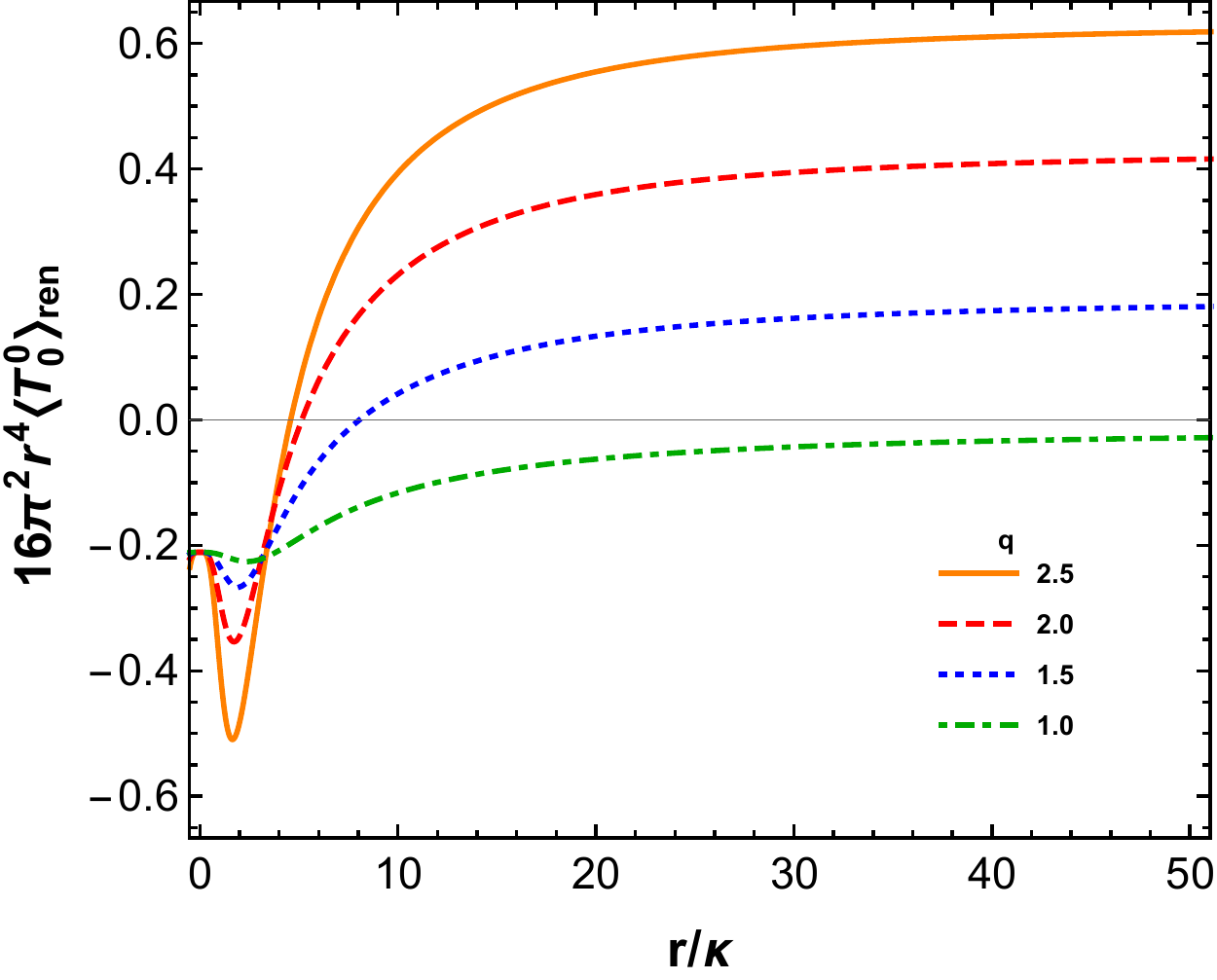}
%
\includegraphics[width=0.41\textwidth]{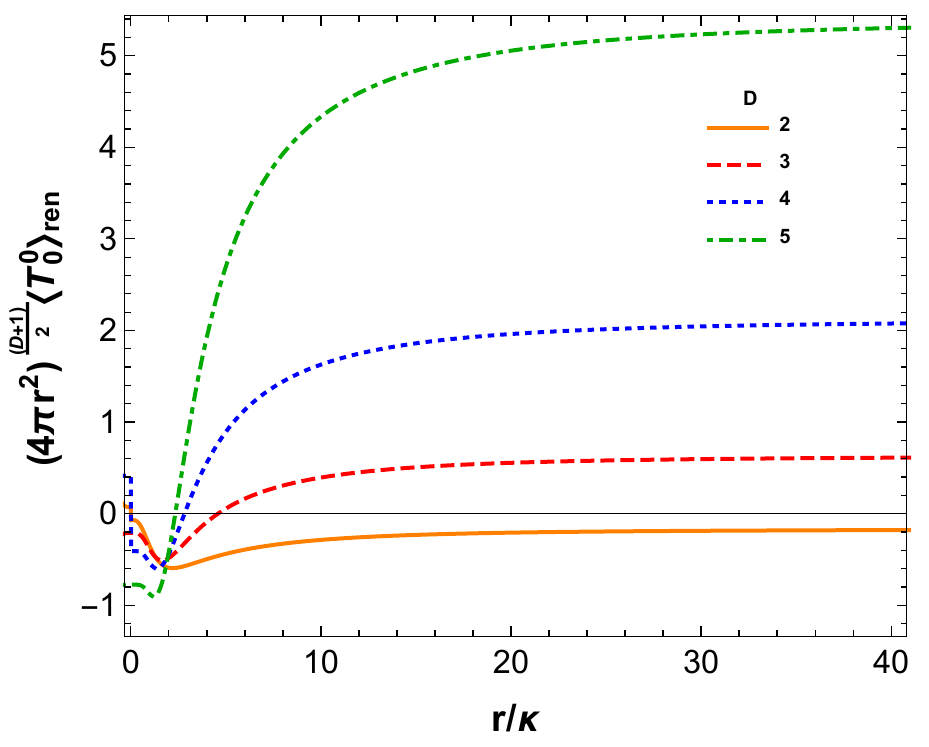}
\caption{\small{On the left: for $D=3$, the (0,0)-component of the energy-momentum tensor, $\left\langle T^{0}_{0}\right\rangle_{{\rm ren}}$, multiplied by $16\pi^2 r^4$, in the massless scalar field case, is plotted as a function of $r/\kappa$ for different values of the cosmic string parameter, $q$. On the right: the energy-momentum tensor, $\left\langle T^{0}_{0}\right\rangle_{{\rm ren}}$, multiplied by $(4\pi r^2)^{\frac{(D+1)}{2}}$, in the massless scalar field case, is plotted as a function of $r/\kappa$ for different values of the spatial dimensions $D$, considering $q=2.5$. Both plots are for $\xi=0$.}} 
\label{f4}
\end{center}
\end{figure}
%
%
\begin{figure}[!htb]
\begin{center}
\includegraphics[width=0.4001\textwidth]{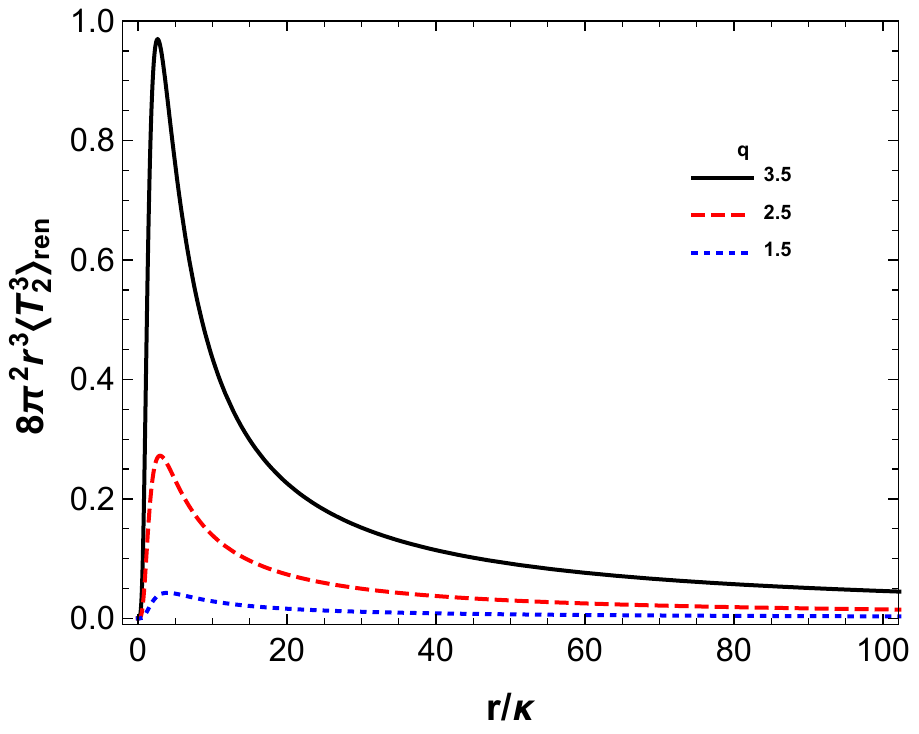}
%
\includegraphics[width=0.4\textwidth]{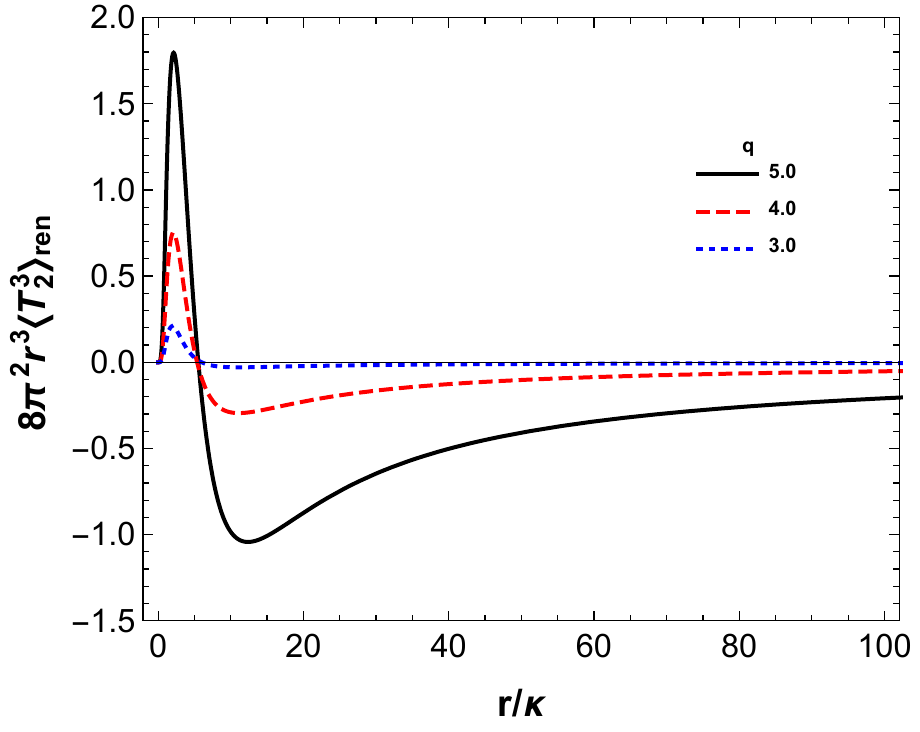}
\caption{\small{For $D=3$, the off-diagonal component of the energy-momentum tensor, $\left\langle T^{3}_{2}\right\rangle_{{\rm ren}}$, multiplied by $8\pi^2r^3$, in the massless scalar field case, is plotted as a function of $r/\kappa$ for different values of $q$.}}
\label{f5}
\end{center}
\end{figure}
%
where $p(y,n,q)$ is given by Eq. \eqref{funp}. As in the massive case, the boost invariance in the massless case of $\left\langle T^{\mu}_{\nu}\right\rangle_{{\rm ren}}$ along the $z$-direction is also broken by the presence of the torsion but it is preserved for the components with $i=4,...,D$. Moreover, we can promptly check that for the massless scalar field, conformally coupled to gravity, the VEV of the energy-momentum tensor (\ref{EMT3}) becomes traceless. One should note that in the limit $\kappa\rightarrow 0$ we recover from Eqs. (\ref{EMT2}) and (\ref{EMT3}) the expressions for the energy-momentum tensor, in the massive and massless scalar fields cases, in the pure cosmic string spacetime \cite{BezerradeMello:2011sm,BezerradeMello:2011nv}. This can essentially be done by using the relations (\ref{B5}) and (\ref{B6}) after taking $\kappa =0$. On the other hand, by considering only the pure screw dislocation spacetime, that is, for $q=1$, the first term on the r.h.s of each Eqs. (\ref{EMT2}) and (\ref{EMT3}) is absent.  As we have already pointed out, the exact closed expressions \eqref{EMT3}-\eqref{EMT3cr} are new results. Below, we will show that in the limit $\epsilon\ll 1$ our results are consistent with the analysis developed in Ref. \cite{DeLorenci:2002jv}.

In special, by writing the $(0,0)$-component of (\ref{EMT3}), for $D=3$, in terms of derivatives with respect to $\epsilon^2$ of the function $g(y,\epsilon,q)$, defined in (\ref{func1}), the following closed expression for the energy density for the massless scalar field is provided.
\begin{eqnarray}
\left\langle T^{0}_{0}\right\rangle_{{\rm ren}}  &=&\frac{1}{\pi^{2}\bar{\kappa}^{4}}\left\{\sideset{}{'}\sum_{l=1}^{[q/2]}\frac{H^{0}_{0,l}(\epsilon,s_l)}{\left[\epsilon^2s_l^2+l^2\right]^3}\right.\nonumber\\
&-&\left.\frac{q}{\pi^2}\int_{0}^{\infty}dy\left[2\epsilon^2\partial^2_{\epsilon^2} + \frac{\left(2\cosh^2(y) + 1\right)}{\cosh^2(y)}\partial_{\epsilon^2}\right]g(y,\epsilon,q)\right\}.
\label{EMT3D}
\end{eqnarray}
Note that all but the off-diagonal component of the energy-momentum tensor can be written in terms of the function, $g(y,\epsilon,q)$. For the off-diagonal case it is possible to perform the sum in $n$ but the resulting expression is too long, so we shall not write it here.

In Fig.\ref{f4}, on the left, we have plotted for $D=3$ the $(0,0)$-component of the energy-momentum tensor multiplied by  $16\pi^2 r^4$, with respect to $r/\kappa$ for different values of $q$. On the right, the plot shows the energy-momentum tensor, multiplied by $(4\pi r^2)^{\frac{(D+1)}{2}}$, in the massless scalar field case, in terms of $r/\kappa$ for different values of the spatial dimensions $D$, considering $q=2.5$. Furthermore, Fig.\ref{f5} shows, for $D=3$, the plot of the off-diagonal component $\left\langle T^{3}_{2}\right\rangle_{{\rm ren}}$, multiplied by $8\pi^2r^3$, with respect to $r/\kappa$ for different values of $q$.

Now we are interested in analysing, for $D=3$, the regimes where $\epsilon\gg 1$ and $\epsilon\ll 1$ for the off-diagonal and $(0,0)$ components. For the latter, in the regime $\epsilon\ll 1$, the leading contribution comes from the second term on the r.h.s of Eq. (\ref{EMT3}) for $n=0$. This contribution has already been analysed and is given by the second term on the r.h.s of Eq. (\ref{asym1}). In fact, in Eq. (\ref{asym1}), if we require that $m\bar{\kappa}\gg 1$, we obtain exactly the asymptotic limit $\epsilon\ll 1$ for the massless scalar field case, with the second term on the r.h.s. being the dominant one. On the other hand, in the limit $\epsilon\gg 1$, analogously to the massive case, the leader contribution is given by the energy density in $(3+1)$-dimensional cosmic string spacetime \cite{DeLorenci:2002jv,BezerradeMello:2011nv, BezerradeMello:2011sm}. This contribution is given by 
\begin{eqnarray}
\left\langle T^{0}_{0}\right\rangle_{{\rm ren}}  &=&\frac{1}{16\pi^2r^4}\left[\sideset{}{'}\sum_{l=1}^{[q/2]}\frac{H^{0}_{0}(s_l)}{s_l^4}\right.\nonumber\\
&-&\left.\frac{q\sin(q\pi)}{\pi}\int_{0}^{\infty}dy\frac{\cosh^{-4}(y)H^{0}_{0}(\cosh y)}{\cosh(2qy) - \cos(q\pi)}\right],
\label{EMT3Dapp2}
\end{eqnarray}
where
\begin{eqnarray}
H_{0}^{0}(v) = 2v^2 - 1.
\label{tt3}
\end{eqnarray}
The sub-leading contribution in this case is found to be
\begin{eqnarray}
\left\langle T^{0}_{0}\right\rangle_{{\rm sub}} &=& \frac{q}{16\pi^2r^4}\int_{0}^{\infty}dy\left[-\frac{9}{4}\cosh^{-3}(y)+\frac{3}{4}\cosh^{-5}(y)\right]\frac{|\kappa|}{r},\nonumber\\
&=&-\frac{27q}{1024\pi r^4}\frac{|\kappa|}{r}.
%
\label{EMT3Dapp3}
\end{eqnarray}
Our leading contribution (\ref{EMT3Dapp2}) coincides with the one obtained in Refs. \cite{DeLorenci:2002jv,DeLorenci:2003wv} in the regime $\epsilon\gg 1$. The sub-leading contribution (\ref{EMT3Dapp3}) falls off with $|\kappa|/r^5$ and is also in accordance with the results of Ref. \cite{DeLorenci:2003wv}. This can be better seen if one takes their function $f(\kappa/r)\propto \frac{r}{\kappa}$ (see the last paragraph of  \cite{DeLorenci:2003wv}). It should be mentioned that, for $q<2$, Refs.  \cite{DeLorenci:2002jv,DeLorenci:2003wv} considered the energy-momentum tensor due to a massless scalar field in the four-dimensional cosmic dispiration spacetime, although the authors did not provid exact closed expressions. In fact, they only considered an analysis for the case $\frac{\kappa}{r}\rightarrow 0$. In this regime, their results is in accordance with ours. The asymptotic behaviours discussed for the massless scalar field can be checked in the plots of Fig.\ref{f4}.

In the regime where $\epsilon\gg 1$, the off-diagonal component becomes 
\begin{eqnarray}
\left\langle T^{3}_{2}\right\rangle_{{\rm ren}}\simeq\frac{1}{8\pi q}\frac{\kappa}{r^4}\left[\sideset{}{'}\sum_{l=1}^{[q/2]}\frac{l\sin(2\pi l/q)}{s_l^{6}}-\frac{q^2}{\pi^3}\int_{0}^{\infty}dy\frac{y\sinh(y)b(y,q)}{\cosh^6(y)}\right],
\label{asymmassless1}
\end{eqnarray}
where the function $b(q,y)$ has been defined in Eq. \eqref{asym3}. The approximation above clearly shows that the off-diagonal component, for the massless case, goes to zero with $\kappa/r^4$, when $r\gg\bar{\kappa}$.

Regarding the regime $\epsilon\ll 1$, we can approximate the off-diagonal component as
\begin{eqnarray}
\left\langle T^{3}_{2}\right\rangle_{{\rm ren}}  &=&\frac{\epsilon^5}{8\pi^2r^3}\bar{b}(\epsilon,q),
\label{asymmassless2}
\end{eqnarray}
where
\begin{eqnarray}
\bar{b}(\epsilon,q) &\simeq&\sideset{}{'}\sum_{l=1}^{[q/2]}\frac{\sin(2\pi l/q)}{l^5}\nonumber\\
&-&\frac{q^2}{\pi^3}\sum_{n=-\infty}^{\infty}\int_{0}^{\infty}dy\frac{y\sinh(2y)p(y,n,q)}{\left[\epsilon^2\cosh^2(y)+n^2\right]^{3}}.
\label{fung}
\end{eqnarray}
For values of $\epsilon$ much smaller than unity the function defined above is positive for the non-integer values of $q>2$ considered in the plot of Fig. \ref{f5}. For $q=1.5$ the first term on the r.h.s is absent and the off-diagonal component also goes to zero from above. This asymptotic behaviour is shown in Fig.\ref{f5}. On the other hand, for integer values of $q>2$ considered in the plot, the off-diagonal component  (\ref{asymmassless2}) goes to zero from below. 

Note that all the numerical analysis of the expressions \eqref{EMT2}-\eqref{arg} and \eqref{EMT3}-\eqref{EMT3cr}, respectively, for the energy-momentum tensor in the massive and massless scalar field cases, as well as all the resulting plots, which are in agreement with the analytical expressions obtained, have never been presented before. Therefore, we have given a detailed investigation of the problem of the Casimir effect of a scalar field in the higher-dimensional cosmic dispiration spacetime.
%
\section{Summary and Discussion}
\label{conc}
%
\hspace{0.55cm}We have considered quantum vacuum fluctuations effects that stem from the nontrivial topology of a spacetime known as cosmic dispiration whose geometry is given by Eq. (\ref{Me2}), and which is a solution of the Einstein-Cartan equations. After obtaining  the exact general solution (\ref{SS3}) of the Klein-Gordon equation in the cosmic dispiration spacetime we were able to calculate closed expressions for the Wightman function for the massive, Eq. (\ref{P32}), and massless, Eq. (\ref{P42}), scalar fields and, as a consequence, the renormalized VEV's of the field squared in both cases were also obtained, Eqs.  (\ref{renVEV2}) and  (\ref{renFS}). These expressions made possible to calculate the VEV of the energy-momentum tensor presented in Eqs.  (\ref{EMT2}), (\ref{EMT2cr}), (\ref{EMT3}) and (\ref{EMT3cr}) for the massive and massless scalar fields, respectively. In particular, in the four-dimensional cosmic dispiration spacetime, our expression for VEV of the energy-momentum tensor in the massless scalar field case shows agreement with the results obtained by the authors in Ref.  \cite{DeLorenci:2002jv,DeLorenci:2003wv}, in the regime  $\frac{r}{\kappa}\rightarrow\infty$ and for $1<q<2$. Nevertheless, we would like to emphasize that our results are more complete and general in the sense we have provided closed expressions for all quantities and for all possible values of the cosmic string parameter $q$, in the massive and massless scalar fields cases. 

The asymptotic behaviours of the VEV's of the energy-momentum tensor in the massive and massless scalar fields cases were also discussed. These behaviours for the $(0,0)$ and $(3,2)$ components , for $D=3$, are shown in the plots of Figs.\ref{f1}-\ref{f5}, including the minimally and conformally coupled cases. For instance, the asymptotic behaviour when $\frac{r}{\kappa}\gg 1$, which has the cosmic string spacetime as the leading contribution, can be confirmed through the plots in Figs.\ref{f1} and \ref{f4} for the massive and massless scalar fields cases, respectively. The behaviour of the off-diagonal component are present in the plots in Figs.\ref{f3} and \ref{f5}. The plot in Fig.\ref{f2} shows that the behaviour of the $(0,0)$-component of the energy-momentum tensor changes drastically near the cosmic string, in the presence of the torsion, when compared with the pure cosmic string case ($\kappa =0$). In the massless scalar field case, we also provided the closed expression (\ref{EMT3D}) for $\left\langle T^{0}_{0}\right\rangle_{{\rm ren}}$ in terms of the function $g(y,\epsilon,q)$, defined in Eq.  (\ref{func1}). With this expression we obtained the asymptotic behaviours  when $\epsilon\ll 1$ and $\epsilon\gg 1$. In special, Eq. (\ref{EMT3Dapp2}) is shown to be the leading contribution in the cosmic string spacetime, with the sub-leading contribution given by Eq.  (\ref{EMT3Dapp3}), which agrees with the one obtained in Refs.  \cite{DeLorenci:2002jv,DeLorenci:2003wv}. The asymptotic behaviours for the off-diagonal component, in the massless case, has also been obtained and are given by Eqs. \eqref{asymmassless1} and \eqref{asymmassless2} when $\epsilon\gg 1$ and $\epsilon\ll 1$, respectively.

It was shown in Ref. \cite{Galtsov:1993ne} that the effect, on a quantized scalar field, of the chiral space-like cosmic string considered here is to produce a shift on the angular momentum quantum number $n$, which is clear if we re-examine the general solution (\ref{SS3}). The latter shows that the shift is given by $n\rightarrow nq - \kappa\nu$ and, as also pointed out by the authors in \cite{Galtsov:1993ne}, is equivalent to the shift produced by the coupling of a given gauge field with a charged scalar field, characterizing the famous Aharonov-Bohm effect. Thereby, in our case, the role of the charge is played by the longitudinal quantum number $\nu$ and the magnetic flux is proportional to the screw dislocation parameter, or more precisely it is given by $2\pi\kappa$. In this sense, it is possible to calculate an induced current density that stem from quantum vacuum fluctuations produced by the cosmic dispiration spacetime. As it was also pointed out in \cite{DeLorenci:2002jv}, this induced current density is due to a flux of longitudinal linear momentum $\nu$. The closed expression for the Wightman function (\ref{P32}) obtained in the present paper can be used for this purpose. This work is in current preparation.

Another possible application of our results is related to the semi-classical back-reaction approach. This approach consists on solving the Einstein equations using as source the energy-momentum tensor associated witht the classical matter plus the $\langle T^{\mu\nu}\rangle_{\mathrm{ren}}$. This important topic opens a new line of investigation. It is our interest to consider the results found for the renormalized VEV of the energy-momentum tensor, and analyze their influence on a semi-classical theory of gravity.

One should emphasize that in a more complete approach a nontrivial structure to the cosmic string core should be taken into account \cite{Gott:1984ef, Hiscock:1985uc, Allen:1990mm, Iellici:1997ud} as to smooth out the singularity in the cosmic string axis. For this matter, two different models have been adopted to describe the geometry inside the string. 
The ballpoint pen model proposed by Gott and Hiscock \cite{Gott:1984ef, Hiscock:1985uc}, for instance, replaces the conical singularity at the string axis by a constant curvature space in the interior region while the flower-pot model \cite{Allen:1990mm} presents the curvature concentrated on a ring, with the spacetime inside the string being flat. In this case the metric \eqref{Me} (for $\kappa =0$) is valid only for the region outside the string. On the other hand, in the Gott and Hiscock model, the manifold can be covered by two maps. The first map covers the interior region of the string, and the second covers its exterior. The metrics are $C^1$-matched on the surface of the string, and there is no surface stress-energy. The Klein-Gordon equation for both regions can be solved exactly. In fact the solutions of the radial equation outside the string is a linear combination of the Bessel and Neumann functions. For the region inside the string, the solution is a Legendre function \cite{Khusnutdinov:1998tf}. These sets of solutions must be continuous, with their first radial derivatives also continuous on the surface of the string. 

Considering a more complete analysis by adopting either model, the VEV associated with the physical observable in the region outside the string is composed by two contributions: the first one, associated with the Bessel function, coincides with our results for $\kappa =0$ (no torsion), and the second contribution is induced by the finite radius of the string, going to zero when we take the limit of its structure going to zero. 

We should also mention that, to the best of our knowledge, none of the models described above have been considering in a unified frame with a screw dislocation and, thus, it is our intention in a future work to study a more realistic model taking into consideration these scenarios. In the present investigation, however, we have neglected the inner structure of the string, that is, we have only considered the effects of the conical structure of the cosmic string as well as the effects of the torsion. A more simplified analysis, neglecting the inner structure of the cosmic string, is the standard procedure to adopt as a first step as it has been done by the authors in Refs. \cite{DeLorenci:2002jv,DeLorenci:2003wv} and by many others in the context of the pure infinitely long and thin cosmic string scenario (see for instance \cite{PhysRevD.35.536, guim1994, LB, Moreira1995365, PhysRevD.35.3779}).

%
\section*{Acknowledgments}
We are grateful to Vitorio De Lorenci and Edisom Moreira for careful reading of the manuscript. We also would like to thank the Brazilian agency CNPq (Conselho Nacional de Desenvolvimento Cient\'{i}fico e Tecnol\'{o}gico - Brazil) for financial support. H.F.M is funded through the research project n$\textsuperscript{\underline{o}}$ 402056/2014-0. E.R.B.M is partially supported through the research project n$\textsuperscript{\underline{o}}$ 313137/2014-5. K.B is partially supported through the research project n$\textsuperscript{\underline{o}}$ 301385/2016-5.
\appendix
\section{Calculation of the Wightman function $W(x,x')$ }
\label{appA}
%
In order to calculate the Wightman function  (\ref{WF}), we need to consider the general solution (\ref{SS3}). This allows us to get
\begin{eqnarray}
W(x,x')=\frac{q}{2(2\pi)^{D-1}}\sum_k \frac{e^{-i\omega_k\Delta t+i{\bf p}\cdot\Delta {\bf r}_{\parallel}}}{\omega_k}e^{i(nq-\kappa\nu)\Delta\varphi+i\nu\Delta Z}\eta J_{|qn-\kappa\nu|}(\eta r)J_{|qn-\kappa\nu|}(\eta r'),
\label{W}
\end{eqnarray}
where $\Delta t=t-t'$, $\Delta {\bf r}_{\parallel}={\bf r}_{\parallel}-{\bf r}'_{\parallel}$, $\Delta\varphi=\varphi-\varphi'$, $\Delta Z=Z-Z'$, $\omega_k^2=m^2+\eta^2+\nu^2+p^2$ and $k=(\eta,n,\nu,p)$ is the set of quantum numbers. Also, we have used the compact notation
\begin{equation}
\sum_k=\int dp^{(D-3)}\int_{-\infty}^{\infty}  d\nu\int_{0}^{\infty}  d\eta\sum_{n=-\infty}^{\infty}. 
 \label{SIapp}
\end{equation}
Moreover, by making a Wick rotation $i\Delta t=\Delta\tau$ in Eq. (\ref{W}) and use the relations \cite{Braganca:2014qma}
\begin{equation}
\frac{e^{-\omega_k\Delta\tau}}{\omega_k}=\frac{2}{\sqrt{\pi}}\int_{0}^{\infty}dse^{-s^2\omega_k^2-\frac{\Delta\tau^2}{4s^2}},
\end{equation}
and
\begin{equation}
\int_{0}^{\infty}d\eta\eta J_{|qn-\kappa\nu|}(\eta r)J_{|qn-\kappa\nu|}(\eta r')e^{-s^2\eta^2}=\frac{1}{2s^2}e^{-\frac{(r^2+r'^2)}{4s^2}}I_{|qn-\kappa\nu|}(rr'/(2s^2)),
\end{equation}
Eq. (\ref{W}) can be re-written as
\begin{equation}
W(x,x')=\frac{q\pi^{\frac{(D-4)}{2}}}{2(2\pi)^{D-1}}\int_{0}^{\infty}\frac{ds}{s^{D-1}}e^{-s^2m^2-\frac{(\Delta\zeta)^2}{4s^2}}\int_{-\infty}^{\infty}d\nu e^{-s^2\nu^2+i\nu\Delta z}\sum_{n=-\infty}^{\infty}e^{inq\Delta\varphi}I_{|qn-\kappa\nu|}(rr'/(2s^2)),
\label{P}
\end{equation}
where $\Delta z=\Delta Z - \kappa\Delta\varphi$ and $\Delta\zeta^2=\Delta\tau^2+\Delta {\bf r}_{\parallel}^2+r^2+r'^2$. Additionally, by making the change of variable $w=rr'/2s^2$ we get
\begin{equation}
W(x,x')=\frac{q}{2(rr')^\frac{(D-2)}{2}(2\pi)^{\frac{D+2}{2}}}\int_{0}^{\infty}dww^{\frac{D-4}{2}}e^{-\frac{m^2rr'}{2w}-\frac{(\Delta\zeta)^2w}{2rr'}}\mathcal{I}(w, \kappa,q),
\label{P2}
\end{equation}
where
\begin{eqnarray}
\mathcal{I}(w, \kappa,q)=\int_{-\infty}^{\infty}\frac{qdh}{\kappa} e^{-\frac{q^2h^2rr'}{2w\kappa^2} + \frac{iqh}{\kappa}\Delta Z}\mathcal{S}(w,h, q),\;\;\;\;\;\;\;\;\;\;\mathcal{S}(w,h, q)=\sum_{n=-\infty}^{\infty}e^{iq(n-h)\Delta\varphi}I_{q|n-h|}(w),\nonumber\\
\label{SF}
\end{eqnarray}
with  $h=\kappa\nu/q$. 

The next step is to work out an expression for the function $\mathcal{I}(w, \kappa,q)$. Let us then start with the sum in $n$ in Eq. (\ref{SF}), i.e,
\begin{equation}
\mathcal{S}(w,h, q)=\sum_{n=-\infty}^{\infty}e^{iq(n-h)\Delta\varphi}I_{q|n-h|}(w).
\label{A1}
\end{equation}
To develop the sum above we can make use of the very useful integral representation \cite{Abramowitz}
\begin{eqnarray}
I_{\beta_n}(w)=\frac{1}{\pi}\int_0^{\pi}dy\cos\left(\beta_ny\right)e^{w\cos y}-\frac{\sin(\pi\beta_n)}{\pi}\int_{0}^{\infty}
dye^{-w\cosh y-\beta_ny},
\label{A2}
\end{eqnarray}
where in our case $\beta_n=q|n-h|$. Thereby, substituting Eq. (\ref{A2}) into Eq. (\ref{A1}), and using the identity  
\begin{eqnarray}
\sum_{n=-\infty}^{\infty}e^{ibn} = 2\pi \sum_{n=-\infty}^{\infty}\delta(b - 2\pi n),
\label{A3}
\end{eqnarray}
the first term on the right-hand side of  Eq. (\ref{A2}) becomes
\begin{eqnarray}
\frac{1}{\pi}\sum_{n=-\infty}^{\infty}e^{iq(n-h)\Delta\varphi}\int_0^{\pi}dy\cos\left(\beta_ny\right)e^{w\cos y}=\frac{1}{q}\sum_le^{-i2\pi lh }e^{w\cos\left(2\pi l/q - \Delta\varphi\right)},
\label{A4}
\end{eqnarray}
with the discrete index $l$ obeying the condition
\begin{eqnarray}
-\frac{q}{2} + \frac{\Delta\varphi}{\varphi_0}\leq l \leq \frac{q}{2} + \frac{\Delta\varphi}{\varphi_0},
\label{A5}
\end{eqnarray}
with $\varphi_0=2\pi/q$. If $\pm$ $\frac{q}{2} + \frac{\Delta\varphi}{\varphi_0}$ are integers then the corresponding terms in the sum on the r.h.s of Eq. (\ref{A4}) should be taken with the coefficient 1/2. Hence, with the result in (\ref{A4}) we find for Eq. (\ref{A1}) that
\begin{eqnarray}
\mathcal{S}(w,h, q)=\frac{1}{q}\sum_le^{-i2\pi lh }e^{w\cos\left(2\pi l/q - \Delta\varphi\right)}-\frac{1}{\pi}\int_{0}^{\infty}
dye^{-w\cosh y}\sum_{n=-\infty}^{\infty}e^{iq(n-h)\Delta\varphi}\sin(\pi\beta_n)e^{-\beta_ny}.\nonumber\\
\label{A6}
\end{eqnarray}
Upon using Eq. (\ref{A6}) to calculate the integral in Eq. (\ref{SF}) we get, for the first term on the right-hand side of Eq. (\ref{A6}), the expression
\begin{eqnarray}
&&\frac{1}{\kappa}\sum_le^{w\cos\left(2\pi l/q - \Delta\varphi\right)}\int_{-\infty}^{\infty}dhe^{-\frac{q^2h^2rr'}{2w\kappa^2}}e^{-i2\pi lh + \frac{iqh}{\kappa}\Delta Z}\nonumber\\
&=&\frac{1}{\kappa}\sum_le^{w\cos\left(2\pi l/q - \Delta\varphi\right)}e^{-\frac{w\Delta Z_l^2}{2rr'}}\int_{-\infty}^{\infty}dhe^{-\frac{rr'q^2}{2w\kappa^2}\left(h - i\frac{w\kappa\Delta Z_l}{qrr'}\right)^2}\nonumber\\
&=&\frac{1}{q}\sqrt{\frac{2w\pi}{rr'}}\sum_le^{w\cos\left(2\pi l/q - \Delta\varphi\right)}e^{-\frac{w\Delta Z_l^2}{2rr'}},
\label{A8}
\end{eqnarray}
where $\Delta Z_l = \Delta Z - \bar{\kappa}l$ and $\bar{\kappa}= 2\pi\kappa/q$.

On the other hand, the second term on the right-hand side of Eq. (\ref{A6}) can be worked out as
\begin{eqnarray}
&&\int_{-\infty}^{\infty}\frac{qdh}{\kappa}e^{-\frac{q^2h^2rr'}{2w\kappa^2} + i\frac{qh\Delta Z}{\kappa}}\sum_{n=-\infty}^{\infty}e^{iq(n-h)\Delta\varphi}\sin(\pi\beta_n)e^{-\beta_ny}\nonumber\\
&=&\int_{-\infty}^{\infty}\frac{qdh}{\kappa}e^{-\frac{q^2h^2rr'}{2w\kappa^2} + i\frac{qh\Delta Z}{\kappa}}\int_{-\infty}^{\infty}dxe^{iqx\Delta\varphi}\sum_{n=-\infty}^{\infty}F(n)\sin(q\pi |x|)e^{-|x|y},
\label{A9}
\end{eqnarray}
where
\begin{equation}
F(n) = \delta\left[x - (n -h)\right],
\label{A10}
\end{equation}
and we have made use of the delta function property $\int \delta(z-z_0)g(z) = g(z_0)$. By making use of Eq. (\ref{A3}) one can show that the sum in $n$ of $F(n)$ above is given by
\begin{equation}
\sum_{n=-\infty}^{\infty}F(n) = \sum_{n=-\infty}^{\infty}e^{i2\pi n(x+h)}.
\label{A14}
\end{equation}
Thus, the result in Eq. (\ref{A14}) allows us to re-writte Eq. (\ref{A9}) as 
\begin{eqnarray}
&&\sum_{n=-\infty}^{\infty}e^{-\frac{\Delta Z_n^2w}{2rr'}}\int_{-\infty}^{\infty}\frac{qdh}{\kappa}e^{-\frac{q^2rr'}{2w\kappa^2}\left(h - i\frac{\Delta Z_n\kappa w}{qrr'}\right)^2}\int_{-\infty}^{\infty}dxe^{ix\Delta\theta_n}\sin(q\pi |x|)e^{-q|x|y}\nonumber\\
&=&\sqrt{\frac{2w\pi}{rr'}} \sum_{n=-\infty}^{\infty}e^{-\frac{w\Delta Z_n^2}{2rr'}}\left\{\frac{\left(q\pi - \Delta\theta_n\right)}{\left(q\pi - \Delta\theta_n\right)^2 + (qy)^2} + \frac{\left(q\pi + \Delta\theta_n\right)}{\left(q\pi + \Delta\theta_n\right)^2 + (qy)^2}\right\},\nonumber\\
&=&\frac{1}{\pi}\sqrt{\frac{2w\pi}{rr'}} \sum_{n=-\infty}^{\infty}e^{-\frac{w\Delta Z_n^2}{2rr'}}M_{n,q}(\Delta\varphi, y),
\label{A15}
\end{eqnarray}
where $\theta_n = 2\pi n + q\Delta\varphi$, $\Delta Z_n = \Delta Z + \bar{\kappa}n$ and
\begin{eqnarray}
M_{n,q}(\Delta\varphi, y) = \frac{1}{2}\left\{\frac{\left(\frac{q}{2} + \frac{\Delta\varphi}{\varphi_0} + n\right)}{\left(\frac{q}{2} + \frac{\Delta\varphi}{\varphi_0} + n\right)^2 + \left(\frac{y}{\varphi_0}\right)^2} - \frac{\left(-\frac{q}{2} + \frac{\Delta\varphi}{\varphi_0} + n\right)}{\left(-\frac{q}{2} + \frac{\Delta\varphi}{\varphi_0} + n\right)^2 + \left(\frac{y}{\varphi_0}\right)^2}\right\}.
\label{A16}
\end{eqnarray}
Therefore, the expression for the function $\mathcal{I}(w, \kappa,q)$ in (\ref{SF}) turns into
\begin{eqnarray}
\mathcal{I}(w, \kappa,q)=\sqrt{\frac{2w\pi}{rr'}}\left[\frac{1}{q}\sum_le^{w\left(\cos\left(2\pi n/q - \Delta\varphi\right) - \frac{\Delta Z_{l}^2}{2rr'}\right)}\right.\nonumber\\
\left.-\frac{1}{\pi^2}\sum_{n=-\infty}^{\infty}\int_{0}^{\infty}dye^{-w\left(\cosh y+\frac{\Delta Z_n^2}{2rr'}\right)}M_{n,q}(\Delta\varphi, y)\right].
\label{A17}
\end{eqnarray}
So, with the expression (\ref{A17}), the Wightman  function (\ref{P2}) is given by
\begin{eqnarray}
W(x,x')&=&\frac{m^{(D-1)}}{(2\pi)^{\frac{(D+1)}{2}}}\left[\sum_{l} f_{\frac{D-1}{2}}\left(m\sigma_l\right)\right.\nonumber\\
&-&\left.\frac{q}{\pi^2}\sum_{n=-\infty}^{\infty}\int_{0}^{\infty}dy f_{\frac{D-1}{2}}\left(m\sigma_{y,n}\right)M_{n,q}(\Delta\varphi,y)\right],
\label{P3}
\end{eqnarray}
where the function $f_{\mu}(x)$ is defined in terms of the Macdonald function, $K_{\mu}(x)$, i.e.,
\begin{equation}
f_{\mu}(x)=\frac{K_{\mu}(x)}{x^{\mu}},
\label{No}
\end{equation}
and
\begin{eqnarray}
\sigma_l^2 &=& \left[\Delta\zeta^2 - 2rr'\cos\left(2\pi l/q - \Delta\varphi\right) + \left(\Delta Z - \bar{\kappa}l\right)^2\right],\nonumber\\
\sigma_{y,n}^2 &=& \left[\Delta\zeta^2 + 2rr'\cosh(y) + \left(\Delta Z + \bar{\kappa}n\right)^2\right].
\label{GD}
\end{eqnarray}

On the other hand, by making $m=0$ in (\ref{P2}) and using (\ref{A17}), the Wightman function for the massless case is found to be
\begin{eqnarray}
W(x,x')&=&\frac{2^{\frac{(D-1)}{2}}\Gamma\left(\frac{D-1}{2}\right)}{2(2\pi)^{\frac{(D+1)}{2}}}\left[\sum_{l} \frac{1}{\sigma_l^{(D-1)}}\right.\nonumber\\
&-&\left.\frac{q}{\pi^2}\sum_{n=-\infty}^{\infty}\int_{0}^{\infty}dy\frac{1}{\sigma_{y,n}^{(D-1)}}M_{n,q}(\Delta\varphi,y)\right].
\label{P4}
\end{eqnarray}
Note that this expression can also be obtained by taking the limit $m\rightarrow 0$ in Eq.  (\ref{P3}).
%
\section{Summation formula for $M_{n,q}(\Delta\varphi,y)$}
\label{appB}
We are now interested in finding an expression for the summation in $n$ of the function $M_{n,q}(\Delta\varphi,y)$. This can be done by using the psi function, $\psi(z)$, and its properties \cite{hille1927logarithmic, Abramowitz}. Thus, let us consider the following definition:
\begin{eqnarray}
\psi(x+iy)=R(x,y)+iI(x,y),
\label{B1}
\end{eqnarray}
where
\begin{eqnarray}
R(x,y)&=&-\gamma+\sum_{n=0}^{\infty}\left[\frac{1}{n+1}-\frac{(n+x)}{(n+x)^2+y^2}\right],\nonumber\\
I(x,y)&=&y\sum_{n=0}^{\infty}\frac{1}{(n+x)^2+y^2},
\label{B2}
\end{eqnarray}
with $\gamma\simeq 0.58$ being the Euler number. Using Eq. (\ref{B2}) (see Ref. \cite{hille1927logarithmic}) it is easy to verify that
\begin{eqnarray}
R(1-x,y)-R(x,y)&=&\sum_{n=-\infty}^{\infty}\frac{(n+x)}{(n+x)^2+y^2}\nonumber\\
&=&\pi\cot(\pi x)\frac{\coth^2(\pi y)-1}{\cot^2(\pi x)+\coth^2(\pi y)}.
\label{B3}
\end{eqnarray}
Therefore, from Eq. (\ref{B3}), we get the final relation  
\begin{eqnarray}
\sum_{n=-\infty}^{\infty}\frac{(n+x)}{(n+x)^2+y^2}&=&\pi\frac{\sin(2\pi x)}{\cosh(2\pi y)-\cos(2\pi x)}.
\label{B4}
\end{eqnarray}
The identity (\ref{B4}) is very important since it allows us to recover, from Eq. (\ref{A17}), the summation formula for the Bessel function $I_{qn}(w)$ in the limiting case $\kappa=0$. This is the case where one considers the cosmic string spacetime only \cite{BezerradeMello:2011sm,BezerradeMello:2011nv}.  

For $\kappa =0$, the sum in $n$ in Eq. (\ref{A17}) reduces to
\begin{eqnarray}
M_q(\Delta\varphi, y) &=& \sum_{n=-\infty}^{\infty}M_{n,q}(\Delta\varphi, y),\nonumber\\
&=& \frac{1}{2}\sum_{j=+,-} \sum_{n=-\infty}^{\infty}\frac{\left(\frac{q}{2} + \frac{j\Delta\varphi}{\varphi_0} + n\right)}{\left(\frac{q}{2} + \frac{j\Delta\varphi}{\varphi_0} + n\right)^2 + \left(\frac{y}{\varphi_0}\right)^2}.
\label{B5}
\end{eqnarray}

In order to apply Eq. (\ref{B4}) for our case, let us make the change $x\rightarrow (q/2 + j\Delta\varphi/\varphi_0)$ and $y\rightarrow y/\varphi_0$. This provides
\begin{eqnarray}
M_{q}(\Delta\varphi, y) =  \frac{ \pi}{2}\sum_{j=+,-}\frac{\sin\left(q\pi + jq\Delta\varphi\right)}{\cosh(q y)-\cos\left(q\pi + jq\Delta\varphi\right)}.
\label{B6}
\end{eqnarray}
In addition, by taking $\kappa=0$ in Eq. (\ref{A17}) and using Eq. (\ref{B6}) we find 
\begin{eqnarray}
\mathcal{I}(w, q)&=&\frac{\sqrt{2w\pi}}{r}e^{- \frac{w\Delta z^2}{2r^2}}\left[\frac{1}{q}\sum_ne^{w\cos\left(2\pi n/q - \Delta\varphi\right)}\right.\nonumber\\
&-&\left.\frac{1}{2\pi}\sum_{j=+,-}\int_{0}^{\infty}dy\frac{\sin\left(q\pi + jq\Delta\varphi\right)}{\cosh(q y)-\cos\left(q\pi + jq\Delta\varphi\right)}e^{-w\cosh y}\right],
\label{B7}
\end{eqnarray}
which is, up to the factor $(2w\pi/r^2)^{\frac{1}{2}}e^{- \frac{w\Delta z^2}{2r^2}}$ that comes from the integral in Eq. (\ref{SF}), the summation formula for Eq. (\ref{A1}) in the case $\kappa=0$, when considering the Casimir energy density in the cosmic string spacetime \cite{BezerradeMello:2011sm, BezerradeMello:2011nv}.
%
%


\end{document}